\documentclass[aps, prd, amsmath, floats, floatfix, onecolumn, nofootinbib]{revtex4-2}
\usepackage{graphicx}
\usepackage{color}
\usepackage{latexsym}
\usepackage{slashed}

\usepackage{amssymb}

\usepackage[hidelinks]{hyperref}%%

\newcommand{\be}{\begin{eqnarray}}
\newcommand{\ee}{\end{eqnarray}}

\newcommand{\bea}{\begin{eqnarray}}
\newcommand{\eea}{\end{eqnarray}}

\newcommand{\pa}{\partial}
\newcommand{\beq}{\begin{equation}}
\newcommand{\eeq}{\end{equation}}
\renewcommand{\ln}{\,\mbox{log}\,}

\newcommand{\eq}[1]{(\ref{#1})}

\renewcommand{\thesection}{\arabic{section}}

\numberwithin{equation}{section}

\let\oldsection\section% Store \section
\renewcommand{\section}{
  \renewcommand{\theequation}{\thesection.\arabic{equation}}
  \oldsection}
\newcommand\ga{\gamma}

\newcommand\ka{\kappa}

\newcommand\La{\Lambda}

\newcommand\si{\sigma}

\newcommand\ph{\varphi}
\newcommand\om{\omega}

\newcommand\na{\nabla}
\renewcommand\pa{\partial}

\begin{document}

%%%%%%%%%%%%%%%%%%%%%%%%%%%%%%%%%%%%%%%%%%%%%%%%%%%%%%%%%%%%%%%%%%%
%%%%%%%%%%%%%%%%%%%%%%%%%%%%%%%%%%%%%%%%%%%%%%%%%%%%%%%%%%%%%%%%%%%

\title{Stringballs and Planckballs for Dark Matter}

\author{Zhongyou Mo}
\email[]{11930796@mail.sustech.edu.cn}

\author{Tib\'{e}rio de Paula Netto}
\email[]{tiberio@sustech.edu.cn}

\author{Nicol\`{o} Burzill\`{a}}
\email[]{nburzilla@outlook.it}
%\email[]{11930956@mail.sustech.edu.cn}

\author{Leonardo Modesto}
\email[]{lmodesto@sustech.edu.cn}

\affiliation{Department of Physics, Southern University of Science and Technology, Shenzhen, 518055, China}

%%%%%%%%%%%%%%%%%%%%%%%%%%%%%%%%%%%%%%%%%%%%%%%%%%%%%%%%%%%%%%%%%%%
%%%%%%%%%%%%%%%%%%%%%%%%%%%%%%%%%%%%%%%%%%%%%%%%%%%%%%%%%%%%%%%%%%%

\begin{abstract}
As a follow up of the seminal work by Guiot, Borquez, Deur, and Werner on ``Graviballs and Dark Matter'',  
we explicitly show that 
in string theory, local and nonlocal higher derivative theories, as well as general asymptotically-free or finite theories,
gravitationally interacting bound states can form when the energy is larger than the Planck energy. 
On the other hand, in higher derivative or nonlocal theories with interaction governed by a dimensionless or a dimensionful coupling constant, the bound states form when the energy is smaller than the Planck energy.    
Such bound states are allowed 
because of the softness of the scattering amplitudes in the ultraviolet region. Indeed, in such theories, the potential is finite while the force is zero or constant in $r=0$. 
Finally, since the bound states that form in the early Universe may have an energy that ranges from the Planck mass to any arbitrarily large or small value, we argue that they can serve as dark matter candidates and/or as the seeds for the structure's formation at large scale in the Cosmos. 
 
\end{abstract}

%%%%%%%%%%%%%%%%%%%%%%%%%%%%%%%%%%%%%%%%%%%%%%%%%%%%%%%%%%%%%%%%%%%
%%%%%%%%%%%%%%%%%%%%%%%%%%%%%%%%%%%%%%%%%%%%%%%%%%%%%%%%%%%%%%%%%%%

\maketitle
\tableofcontents%%
\section{Introduction}

In this work, we extend to several proposals for quantum gravity, the idea introduced for the first time in the seminal paper \cite{Guiot2020}.
 Therefore, it is a must to start with an incomplete list of approaches to quantum gravity. 
In order to get a well-defined theory for quantum gravity in the quantum field theory framework, there are basically five current proposals: String Theory (ST)~\cite{StringTheory, Polchinskibook}, Nonlocal Quantum Gravity (NLQG)~\cite{Krasnikov, Kuzmin, Modesto, ModestoLeslawF, ModestoLeslawR, Modesto:2021okr, Modesto:2021ief, Universally, FiniteGaugeTheory}, Higher Derivative Quantum Gravity (HDQG)~\cite{shapiro3, ShapiroModestoLW, ModestoLW}, Stelle's Quantum Gravity (SQG)~\cite{Stelle}, and Asymptotically Safe Quantum Gravity (ASQG)~\cite{ASQG,ASQG2,ASQG3}. String theory, NLQG, and HDQG are finite theories, but the minimal NLQG and HDQG can be simply super-renormalizable and asymptotically free. ASQG is also somehow finite in the Wilsonian sense, while SQG is the only renormalizable theory of gravity. Therefore, SQG is a unique theory according to the simple renormalizability of all the other fundamental interactions. 

Perturbative unitarity is achieved in all the listed theories above. In ST (expanding around the Minkowski background) and in NLQG (expanding around any solution of Einstein's equations of motion), the spectrum consists only of physical particles (detectable propagating degrees of freedom) and the amplitudes, consistent with the Cutkowsky rules and unitarity~\cite{Briscese:2018oyx, Briscese:2021mob, spallucci}, are obtained by means of analytic continuation of the external energies from purely imaginary to real values. On the other hand, in SQG, the ghost field is removed by hand from the spectrum, and the perturbative unitarity is achieved at any loop order by the mean of the Anselmi-Piva prescription that avoids the ghosts to be created in the loop Feynman diagrams~\cite{Anselmi:2017ygm, AnselmiPiva1, AnselmiPiva2, AnselmiPiva3, AnselmiNoMinkowski}. Finally, in ASQG, unitarity should emerge at non-perturbative level~\cite{Wetterich}. 

That said, we are legitimate to ask the following question: can somehow these theories explain one of the greatest mysteries of gravitational physics, namely dark matter?
Indeed, such an issue is not a recent discovery but actually goes back to the $1933$~\cite{Zwicky:1933gu}, and the most common interpretation relies on the presence of a huge fraction of missing dust matter in our Universe.
Such presence of exotic matter is needed to explain various phenomena: the galactic rotation curves, the structure formation in the early Universe, the cosmological microwave background (CMB) radiation, the gravitational lensing, and the so-called ``bullet cluster''.
However, we must honestly say that this point of view is based on the assumption that the gravitational interaction, as well as all the other fundamental interactions, is the same at every energy scale. In particular, usually, people modify the ``right-hand side'' of Einstein's equations (EE) while assuming correct the ``left-hand side''. Indeed, it is well known that it is very hard to modify consistently the EE at large distances without introducing instabilities and facing off other issues. 
Therefore, in two recent papers \cite{Li:2019ksm,Modesto:2021yyf} we addressed the galactic rotation curves' problem without modifying gravity, but reading the Einstein-Hilbert theory (EH) as Einstein's conformal gravity (ECG) in the Higgs phase after the Weyl conformal symmetry is spontaneously broken. This is tantamount to trying to understand gravity instead of modifying gravity.
In the context of conformal gravity there are also indications that it can be possible to remedy the missing matter in the spectrum of the CMB radiation as suggested in \cite{Mukhanov} (see \cite{Sebastiani:2016ras} for a review on Mimetic Gravity). 
In the latter paper, the authors showed that in Mimetic-Gravity, a theory very similar to ECG, the FRW equation has an extra energy density term, but no new extra propagating degrees of freedom. Such energy density behaves as dust matter and could fix the issue with the CMB spectrum without the need of dark matter 
\cite{Myrzakulov:2015kda, Vagnozzi:2017ilo}. In particular, in the latter papers it was shown that the phenomenology of DM on galactic scales can indeed be explained by minimal well-motivated extensions of mimetic gravity.

So far so good, but what about the bullet cluster and the structures' formation? At the moment, we do not know how to address the former issue, but we can provide the following proposal for the latter one. 
We here closely follow the bright idea suggested in \cite{Guiot2020}, where the authors proposed that dark matter could be made of massless particles (gravitons in \cite{Guiot2020}) confined in a bound state.
In this regard, as the outcome of this paper, we will show that perturbative bound states at high energy can be created in both nonlocal gravity and higher derivative gravity because such theories are asymptotically free or finite at short distances.
Indeed, the short-range weakness of all fundamental interactions described by NLQG and HDQG motivated us in proposing that the formation of structures in the Universe is not due to the presence of extra matter but to bound states of gravitons or other particles of the standard model in its local or nonlocal ultraviolet completion. 
Afterward, such bound states will serve as seeds on which other matter will clump. 
The same proposal also applies to string theory, where the tree-level scattering amplitudes are soft at high energy. 
Notice that in this Universe scenario we only have the five percent of observed matter and no dark matter at all. 
Because of the almost absence of interactions, in the early Universe, even supermassive bound states can form that later will gravitationally attract baryonic matter to form larger structures. Notice that in this scenario the amount of dark matter in the Universe can be neglected, but still we can explain the structures formation. Nevertheless, the conventional dark matter scenario is not excluded. Indeed, we can also assume that the bound states characterize the amount of the needed extra matter, namely the twenty-five percent of the claimed dark matter. This was actually the proposal in \cite{Guiot2020}. 

Summarizing we can have two scenarios: (I) a first one without dark matter in which the structures formation is due to the weakness of all fundamental interactions in the early Universe, (II) a second scenario in which bound states are created in the early Universe to serve as the about twenty-five percent of missing matter in the Cosmos. 
Both the scenarios need bound states that will be the real topic of this paper. 
Indeed, we will show that in string theory and nonlocal or higher derivative theories, perturbative bound states are allowed while the same states are not allowed in Einstein's or general two-derivative theories. 
We will follow the approximation introduced in \cite{Guiot2020} and consider the movement of massless particles in an ``effective potential'' defined starting from the scattering amplitudes in the Regge's limit $t\ll s$. 
Indeed, after simple manipulations, we will be able to infer about the ``effective equations of motion'' (EEoM) regulating the dynamics of massless particles (notice that in the early Universe, we can always forget the mass term in the dispersion relation). 

We would like to stress once again that the content of this paper is very much quantitative and less speculative regardless of the reasons given above, on which perhaps many readers may not agree.
Indeed, we will consider the following explicit examples: 
(I) a sterile-scalar field in local and nonlocal higher derivative gravity, (II) string theory, (III) nonlocal scalar electrodynamics, (IV) a $\phi^3$ higher derivative or nonlocal field theory, and (V) general asymptotically free or finite NLQG and HDQG.

We will name the perturbative bound states of sterile massless scalars ``{\em gravi-scalarballs}'', those made of pions in scalar-electrodynamics by ``{\em electroballs}''. In string theory, we will name the bound states ``{\em stringballs}'', while in general, we will baptize the bound states of any kind of massless particles ``{\em Planckballs}'' on the footprint of the ``{\em graviballs}'' (bound states of gravitons) as stated in \cite{Guiot2020}. 
In what follows we use the metric signature $(-,+,+,+)$ and also adopt natural units $c = \hbar = 1$.

%%%%%%%%%%%%%%%%%%%%%%%%%%%%%%%%%%%%%%%%%%%%%%%%%%%%%%%%%%%%%%%%%%%
%%%%%%%%%%%%%%%%%%%%%%%%%%%%%%%%%%%%%%%%%%%%%%%%%%%%%%%%%%%%%%%%%%%

\section{Effective equations of motion} 
\label{sec2}

In order to investigate the emergence of bound states, we follow Ref.~\cite{Guiot2020} and consider a system of two identical massless particles traveling in the $xy$-plane in opposite directions with the same energy $E_1=E_2=\omega$. According to \cite{Guiot2020}, our computation is ``semi-classical'' because we study the relativistic equations of motion based on the energy potential defined as the Fourier transform of the $t$-channel scattering amplitude $A_t (t)$ in the Regge limit $ t \ll s$ and $t=-q^2 = -\vec{q}\,^2 \rightarrow 0$, namely 
\be 
\label{potential-1}
V(|\vec{r}_1-\vec{r}_2|) =  - \frac{1}{4E_1  E_2} \int \frac{d^{3} q}{(2 \pi)^3 } \, e^{i \vec{q} \cdot ( \vec{r}_1-\vec{r}_2) } 
 A_t (-\vec{q}\,^2 ) \, ,
\ee
where $\vec{r}_i$ denotes the position of the $i$-th particle, and the Mandelstam variables are defined as follows, 
\be
s = - (p_1+p_2)^2, 
\qquad 
t = -( p_1-p_3)^2 = - q^2, 
\qquad 
u = - (p_1 - p_4)^2,
\ee
with $q^2$ being the transfered momentum.
We assume that the initial 3-momenta of the particles are along the $x$ direction, and we choose to work in the center of mass frame, such that we have the initial $4$-momenta:
\be
\label{initalP}
p_{1}^{\,\mu} = (\omega,-\omega,\,0,\,0) 
 \qquad
\text{and} 
 \qquad
p_{2}^{\,\mu} = (\omega,\omega,\,0,\,0) \, .
\ee 
We here deal with a central force $\vec{F}_i = -\vec{\nabla}_{{i}} V $ acting on the $i$-th particle. 
Therefore, using the condition $\vec{F}_1=-\vec{F}_2$ and the symmetries of the system, we can reduce it to a single-particle problem. In particular, the center of mass is at rest because in the initial configuration $\vec{p}_1+\vec{p}_2=0$. 
Then, choosing the origin of the coordinates system such that at the initial time  $\vec{r}_0 \equiv \vec{r}_1(t_0)=-\vec{r}_2(t_0)$, we have $\vec{r}_1=-\vec{r}_2$. 
Moreover, for the same reason for the velocities, we have $\vec{v}_1=-\vec{v}_2$. Thus, we can consider the 
reduced system of a single particle described by the vector: 
\beq
\label{reducedR}
\vec{r} \equiv \frac{\vec{r}_1 - \vec{r}_2}{2} \, 
.
\eeq

The equations of motion describing our system can be derived from the following relativistic relations for massless particles, namely~\cite{Guiot2020}
\begin{align}
\label{EoM11}
& {\vec{v}} \equiv  \dot{\vec{r}}
\, ,
\\
&
\dot{\vec{v}}
= \frac{1}{\omega}[\vec{F}-(\vec{v}\cdot\vec{F})\vec{v}] 
\, ,
\label{EoM1}
\\
&
\label{EoM12}
{\vec{v}} \,^2 = 1 \, ,
\end{align}
where \eq{EoM12} means that the particles travel at the speed of light, 
and the force and the potential are:
\begin{align}
\label{red_force}
&
\vec{F}(\vec{r}) = -\frac{1}{2} \frac{d V(r)}{dr} \frac{\vec{r}}{r} = F(r) \, \frac{\vec{r}}{r} \, ,
%\label{force}
\\
&
\label{red_pot}
V(r) =- \frac{1}{16 \pi^2  \omega^2 r} \int_{0}^{\infty} d q \, q  \sin{(2 q r )} A_t (-\vec{q}\,^2 )
\, ,
\end{align}
with $r = |\vec{r}\,|$ and $q = |\vec{q}\,|$. Throughout this work we will use the ``dot'' notation to denote the differentiation with respect to the time $t$. In deriving Eq.~\eq{red_pot} we used \eq{reducedR} in \eq{potential-1}, performed the integration over the angles, and used the relation $(2E_1)(2 E_2) = 4 \omega^2$ for the normalization factor. The extra factor of $1/2$ in \eq{red_force} comes from the derivative with respect to the reduced coordinate \eq{reducedR}. The initial conditions are parametrized as     
\bea
\vec{r}_0 = \left(\frac{a}{2},\frac{b}{2}\right)
\quad \text{and} \quad
{\vec{v}}_0 = (-1,\,0)
.
\label{EoM2}
\eea

Now we go beyond the analysis of Ref.~\cite{Guiot2020} (which was purely numerical) and carefully study the equations \eq{EoM11}--\eq{EoM1} in order to extract some general analytical constraints about the formation of bound states. It will be convenient to make use of polar coordinates in $xy$-plane, namely 
\beq
\label{polar}
x = r \cos \varphi, 
\qquad \qquad
y = r \sin \varphi. 
\eeq
Then, using the known formulas for the velocity and the acceleration in polar coordinates~\cite{anode} 
\bea
\vec{v} = \dot{r} \, \hat{e}_r + r \dot{\ph} \, \hat{e}_\ph
\, ,
\qquad 
\dot{\vec{v}} = ( \ddot{r} - r \dot{\ph}^2 ) \, \hat{e}_r + (2 \dot{r} \dot{\ph} + r \ddot{\ph}) \, \hat{e}_\ph
\, ,
\eea
and also a central force $\vec{F} = F(r) \, \hat{e}_r$, the radial and angular components of \eqref{EoM1} are respectively given by:
\bea
\ddot{r}-r\dot{\varphi}^2&=&\frac{(r\dot{\varphi})^2 F(r)}{\omega}
\label{dotR}
\, ,
\\
2\dot{r}\dot{\varphi}+r\ddot{\varphi}&=&-\frac{r\dot{r}\dot{\varphi}F(r)}{\omega}
\, .
\label{dotphi}
\eea
In polar coordinates the initial conditions \eq{EoM2} read:
\begin{equation}
\begin{split}
&r(t_0)=\frac{\sqrt{a^2+b^2}}{2}, \qquad \qquad \,\,\,\,\,
\varphi(t_0)=\arcsin\frac{b}{\sqrt{a^2+b^2}}, \\
&\dot{r}(t_0)=-\frac{a}{\sqrt{a^2+b^2}}, \qquad \qquad
\dot{\varphi}(t_0) =\frac{2b}{a^2+b^2}.
\label{initial}
\end{split}
\end{equation}
Consistently with the Regge limit $t \ll s$, the initial angle has to be small, i.e  $\varphi(t_0) \approx 0$. Therefore, we should assume $b \ll a$.

A first integral of motion can be obtained multiplying~\eqref{dotphi} by $r$ and defining the angular momentum by the following relation, 
\beq
\vec{L} = \vec{r} \times \vec{p} \, 
. 
\eeq
Indeed, Eq.~\eqref{dotphi} can be integrated to give:
\begin{equation}
\label{phiL}
\dot{\ph} = \frac{L}{\om r^2}
\, , %\qquad \qq \text{where}
\end{equation}
where
\begin{equation}
L(r) = L_0 \, e^{\frac{V (r) - V(r_0)}{2\omega}} \, ,
\qquad 
L_0 \equiv  L_0(t_0) = \frac{b \, \omega}{2}
\, .
\label{L}
\end{equation}

Instead of solving \eq{dotR}, we consider the simpler equation~\eq{EoM12} in polar coordinates\footnote{Indeed, after plugging Eq.~\eq{phiL} into \eq{dotR} and \eq{conservation} one can show that \eq{dotR} and \eq{conservation} give the same solutions for $r(t)$.},
\be
\dot{r}^2+(r\dot{\varphi})^2=1.
\label{conservation}
\ee
Replacing \eq{phiL} into the above equation \eq{conservation}, we end up with:
\begin{equation}
\begin{split}
& \dot{r} = \pm \sqrt{ 2 \left[ E_\text{\text{eff}} - U(r) \right]}  
\, ,
\qquad
E_\text{\text{eff}} = \frac{1}{2},
%\label{conservation E} %\nonumber 
\\
%\ee
%where
% \be
% U(r) = \frac{1}{2} \left( \frac{L_0}{\omega r} \right)^2 
% e^{\frac{V(r) - V_0}{\omega}}
% .
% \label{effective potential energy}
% \ee
&
U(r) = U_0 \left(\frac{r_0}{r} \right)^2  
e^{\frac{V(r) - V(r_0)}{\omega}}
\, ,
\qquad 
U_0 = U(r_0) = \frac{b^2}{2(a^2+b^2)}
\, .
\label{effective potential energy}
\end{split}
\end{equation}
Therefore, we have reduced the two-dimensional dynamical system \eq{dotR}--\eq{dotphi} to a one-dimensional problem for a particle moving in an ``effective potential" $U(r)$ with the ``effective energy" $E_{\text{eff}} = 1/2$. All possible trajectories with $U(r) \leqslant 1/2$ are allowed.

However, it is important to notice that, despite the function $U(r)$, which is defined in~\eq{effective potential energy}, resembles an effective potential of classical mechanics (and it can be analyzed with in the general framework that we can find   in standard textbooks~\cite{anode}), its physical interpretation is much more subtle for the following reasons. First of all, we notice that $U(r)$ and $E_{\text{eff}}$ 
are dimensionless quantities, and, in particular from \eqref{effective potential energy}, we see that they have the physical meaning of velocities squared rather than energies. Indeed, the Eq.~\eq{effective potential energy} is nothing else but the rewritten equation for the constant modulus of the velocity (equals to the speed of light) of a massless particle that cannot be at rest. For this reason, we notice that even if the interaction potential $V(r)$ is always attractive, i.e., $V(r)<0$ $\forall \, r$, the effective potential $U(r)$ is always a positive-defined function, see Eq.~\eq{effective potential energy}, otherwise \eq{EoM12} would be violated. Secondly, $U(r) \leqslant E_{\rm eff} = 1/2$ is just a reflex of the limiting velocity of a massless particle; if the particles could access regions where $U(r) > 1/2$ it would acquire superluminal velocities. Therefore, any value of $r$ with $U(r) > 1/2 $ is not allowed. 
In particular, even when there is no interaction between the particles, $U(r)$ is non-zero. Using $V(r)=0$ in~\eq{effective potential energy}, we see that the effective potential boils down to a hyperbola
\beq
U^{\text{free}}(r)=\frac{b^2}{8 r^2} \, ,
\eeq 
and substituting this expression into the first line of \eqref{effective potential energy}, after solving for $r$, one can recover the parametric equation of a straight-line with constant velocity $|\vec{v}| = 1$, as expected in the case where the interactions are turned off.

In what follows, we make use of a qualitative analysis of the effective potential $U(r)$ to infer about the existence of bounded solutions. In order to have bound states, the effective potential~\eq{effective potential energy} must have at least one local minimum $U_{\text{min}}$ for a finite value of $r$. Of course, no bound states can be formed whether the impact parameter vanishes, $b = 0$, because we get $U_0 = 0$, and finally $U(r) = 0$. 
Therefore, from now on, we will assume that $b \neq 0$. 

Before to proceed, we need to require some properties for the interaction potential~\eq{red_pot}. As discussed in the introduction, our main goal is to consider modified potentials inspired by several classical and quantum gravity models. However, before discussing such specific examples, we can infer about many general results on the formation of bound states by only looking at general properties of the potential.

Let us assume that the potential $V(r)$ satisfies the following conditions:
\begin{itemize}
\item[(i)] The singularity in $r=0$ is removed in the proposed classical theory or by quantum field theory corrections, namely $V(r)$ is an analytical function 
around $r=0$ such that $\lim\limits_{r \to 0} V(r) < \infty$. 
\item[(ii)] At large distances, the quantum effects are negligible, and the potential reproduces the Newtonian one, i.e., 
\be
V(r) \underset{r \to \infty}{\sim} -\dfrac{1}{r}.
\ee
\end{itemize}
Therefore, if the conditions (i)--(ii) holds, we have from~\eq{effective potential energy} the following results for $U(r)$, 
\be
\lim_{r \to 0} U(r) = \infty \, , 
\qquad  
U(r) \underset{r \to \infty}{\sim} \frac{1}{r^2}
\, .
\label{Uconditions}
\ee
The first condition in \eq{Uconditions} means that we have an infinite potential wall at the origin, so the system can never reach the point $r = 0$.

We now look for extreme points for the effective potential~\eq{red_pot}. Taking the derivative of $U(r)$ with respect to $r$, we get:
\be
\frac{dU}{dr} = \frac{2 U_0 \,r_0^2}{r^3}\,  e^{\frac{V(r)- V(r_0)}{\omega}} \, Q(r)
\, 
,
\label{dW}
\ee
where we introduced the new function:
\be
Q(r) = -1 - \frac{r}{ \omega} F(r) \, .
\label{Q}
\ee
Because of (i)--(ii), the zeros and the change of sign of \eq{dW} are related to the function $Q(r)$, which mainly depends on the force $F(r)$. 

Under the assumptions (i)--(ii) for the potential, we have the following corollaries for the force $F(r)$. 
\begin{itemize}
\item[(iii)] $F(r)$ is also an analytical function around $r=0$, and we have:
\be
\lim\limits_{r \to 0} F(r) < \infty \, .
\ee
\item[(iv)] The behavior of the force at large distances is:
\be
F(r) \underset{r \to \infty}{\sim} -\dfrac{1}{r^2} \, .
\ee

\end{itemize}

\begin{figure}[t]
\includegraphics[scale=0.8]{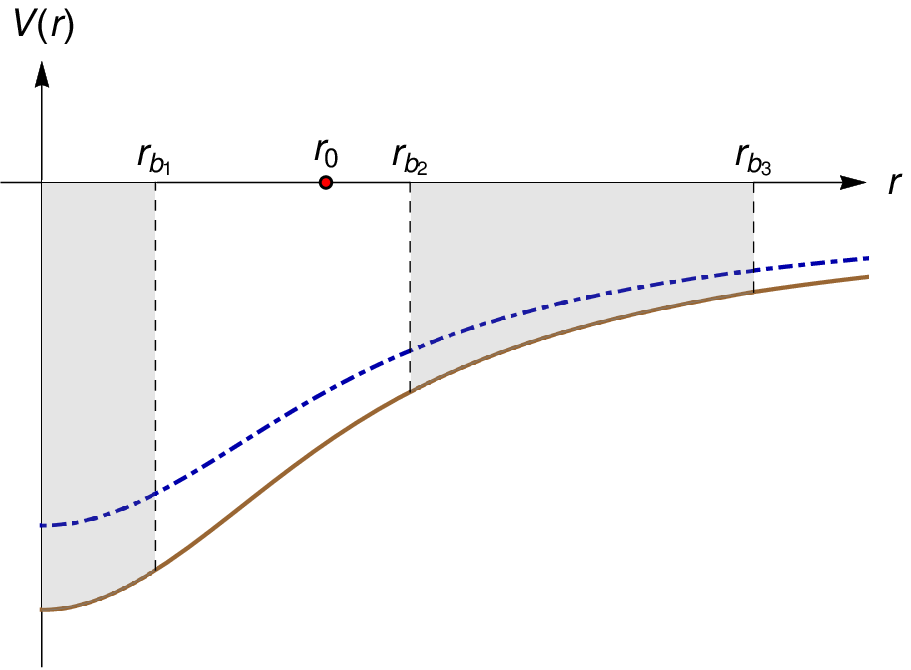}
\includegraphics[scale=0.8]{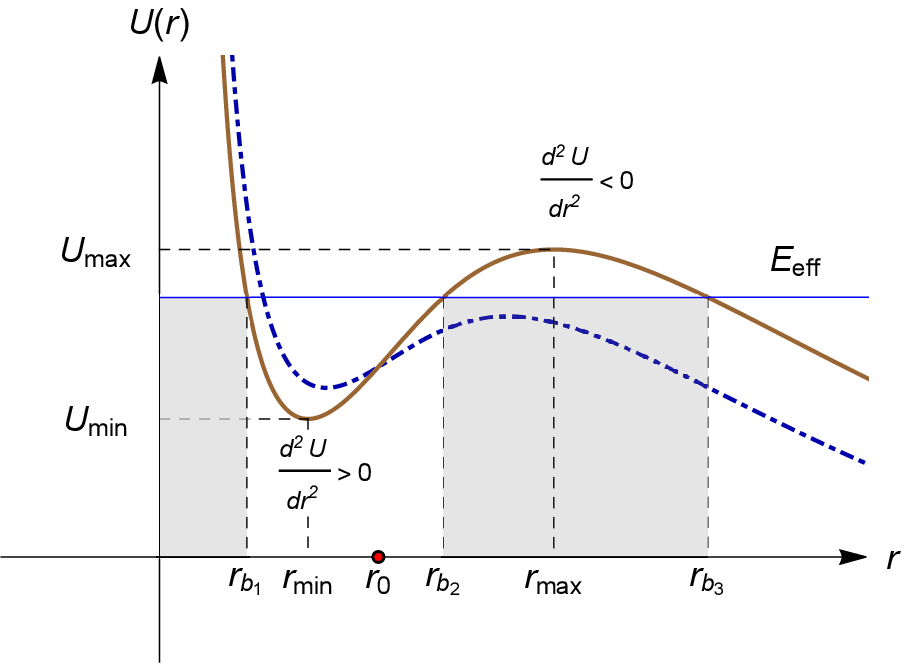}
\includegraphics[scale=0.8]{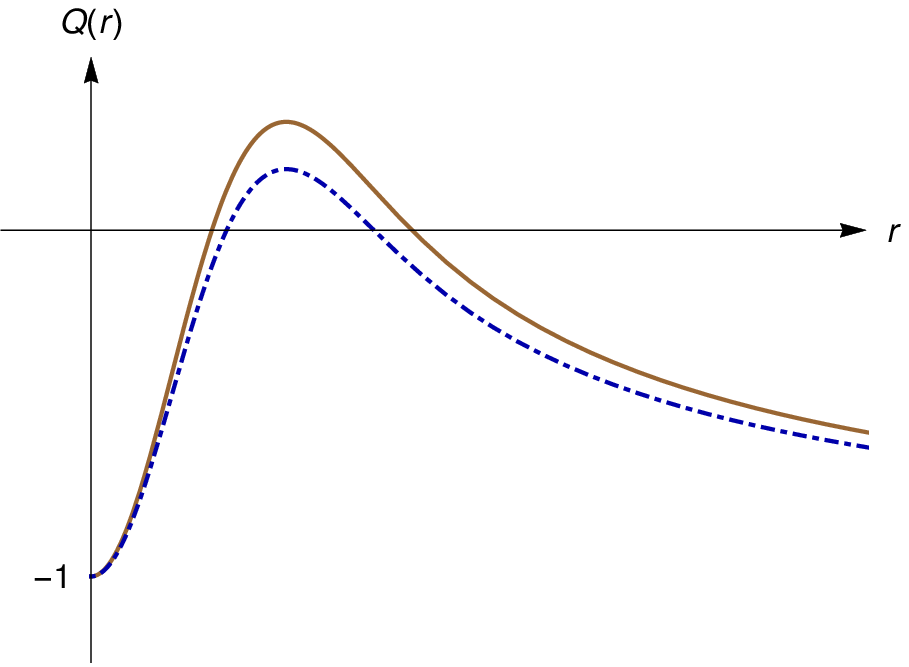}
\caption{
Qualitative plots showing the comparison between $V(r)$, $U(r)$, and $Q(r)$ in the case where the scattering potential $V(r)$ satisfies the requirements (i)--(ii).
The shaded areas represent the forbidden values for the radial coordinate $r(t)$, where $U(r) > 1/2 = E_{\rm eff}$. 
Although $V(r)$ is always negative, the effective potential $U(r)$ is a positive-defined function. The red dot represents the initial condition $r_0$. For the solid-line curves, all the conditions in~\eq{nsconditions} are satisfied. In this case, since $r_{b_1} < r_0 < r_{b_2}$, the bounded solutions are allowed; the time evolution of $r(t)$ is confined in between  $r_{b_1}$ and $r_{b_2}$. Although $U_{\rm \min}$ is only a local minimum, the region $[r_{b_3},+\infty]$ (while in principle not forbidden) remains inaccessible to the system. Indeed, accessing such region (with the initial condition shown in the plot) would constitute a violation of the condition \eqref{EoM12}, which means a velocity greater than the speed of the light. For this reason, the bounded movement is stable, and in order to break the confinement the particle must be a tachyon.
The dot-dashed line represents the case where (I) in~\eq{nsconditions} is satisfied while the condition (II) is violated. Despite the fact that $Q_{\rm max} >0$ because (II) is not true, there exist only a single turning point where $U(r_b) = E_{\rm eff}$. 
Hence, the particles will scatter through the infinity, and bounded solutions are not possible in this situation.
} \label{graphpot}
\end{figure}

The conditions (iii)--(iv) implies that $Q(r)$ is a continuous and differentiable function, and that 
\be
\lim_{r \to 0} Q(r) = \lim_{r \to \infty} Q(r) = - 1
\, .
\label{Qconditions}
\ee
Therefore, Rolle's theorem guarantees that the function~\eq{Q} has at least one critical point $Q'(r_c) = 0$. 
Let us assume that at the critical point $Q(r)$ takes the maximum value $Q_{\text{max}}$. If this maximum is positive, namely 
\beq \label{Qmax}
Q_{\text{max}} > 0 \, , 
\eeq
then the equation $Q(r) = 0$ has at least two solutions and the derivative $dQ/dr$ changes the sign. 
In the simplest case, the effective potential~\eq{effective potential energy} has two equilibrium points~$r_e$ and, since $U(r)$ is continuous on $(0, \infty)$, the conditions~\eq{Uconditions} imply that the effective potential has a local minimum $U_\text{min} = U(r_\text{min})$ and a local maximum $U_\text{max} = U(r_\text{max})$, with $r_\text{max} > r_\text{min}$ and $r_\text{min}\neq 0$. Hence, if $U_{\rm max} > E_{\rm eff}$ there are three solutions of the equation $U(r_b) = E_\text{eff}$, that, together with the condition $U\leqslant 1/2$, define the regions of the phase space that are allowed for the evolution of the system (see Fig.~\ref{graphpot}). As we can see from \eqref{conservation} and \eqref{effective potential energy}, violating the condition $U\leqslant 1/2$ would be equivalent to violate the Eq.~\eqref{EoM12}. The two turning points\footnote{Indeed, it is possible that $r_{b_1} = r_{b_2}$ for $U_0 = U_\text{min} = E_{\rm eff} = 1/2$. However, as \eq{effective potential energy} shows $U_0 = 1/2$ is only possible for $a=0$. In general, any solution with $a=0$ corresponds to a circular motion with radius $r_0 = r_b = b/2$, $U(r_b) = U_0 = E_{\rm eff}$. In this situation, the scattering condition $a \gg b$ is violated. Therefore, we always assume that $r_0 \neq r_b$.} $r_{b1}$ and $r_{b2}$ with $r_{b_1} < r_{\rm min} < r_{b_2}$, define the region of bounded motion $[r_{b1},r_{b2}]$, while in the region $[r_{b3},+\infty]$, the two particles would reach a minimal mutual distance $r_{b3}$ before scattering away to infinity. 
In the case where the initial position is such that $r_{b_1} < r_0 < r_{b_2}$, the movement is confined in the interval $[r_{b1},r_{b2}]$, and bound state solutions are allowed. 
In Fig.~\ref{graphpot}, we have a plot of a typical graph for the potential $V(r)$, the effective potential $U(r)$ and the function $Q(r)$, under the assumptions (i)--(ii) that allow for bound solutions.
Otherwise, if  $Q_{\text{max}} \leqslant 0$, or $Q(r)$ possesses only one local minimum, the function $Q(r)$ can not be positive; therefore, the effective potential is a monotonically decreasing function, and bound states are not allowed.

In summary, the necessary and sufficient conditions for bound states are:
\beq\label{nsconditions}
\boxed{
\text{(I)} \quad  Q_{\text{max}} > 0 \, , \quad
\text{(II)}\quad  U_\text{max} > E_\text{eff}\, , \quad
\text{(III)}\quad  r_{b_1} < r_0 < r_{b_2} \, , \quad
\text{(IV)}\quad  0 < r_{b_1} < r_\text{min}  < r_{b_2} < r_\text{max} \, ,
}
\eeq
where $r_0 = r(t_0)$ was defined in~\eq{initial}. 

In order to find bound states of particles with energy $\omega$, 
in the rest of the paper we will assume $Q_\text{max} >0$.
On the other hand, by studying the change of sign in the derivative $dQ/dr$ we can found bounds on the equilibrium points $r_e = \{r_\text{min},r_{\max}\}$. 
For instance, since $Q(r_e) = 0$, we have
\be
\label{d2W}
\frac{d^2U}{dr^2} \Big|_{r = r_e} =  \frac{2 U_0 \,r_0^2}{r^3_e}\,  e^{\frac{V(r_e)-V(r_0)}{\omega}} \, \frac{dQ}{dr}\Big|_{r = r_e} \, ,
\ee
thus the sign of the second derivative of the effective potential is directly related to the sign of the first derivative of $Q(r)$. Evidently,
\beq
\begin{split}
&
\frac{d^2 U}{d r^2} \Big|_{r = r_e} > 0, \qquad \mbox{for} \qquad r_e = r_\text{min} \, , 
%\qquad \text{and}
\\
&
\frac{d^2 U}{d r^2} \Big|_{r = r_e} < 0, \qquad \mbox{for} \qquad r_e = r_\text{max}
\, .
\label{d2W2}
\end{split}
\eeq 

\begin{figure}[t]
\includegraphics[scale=0.8]{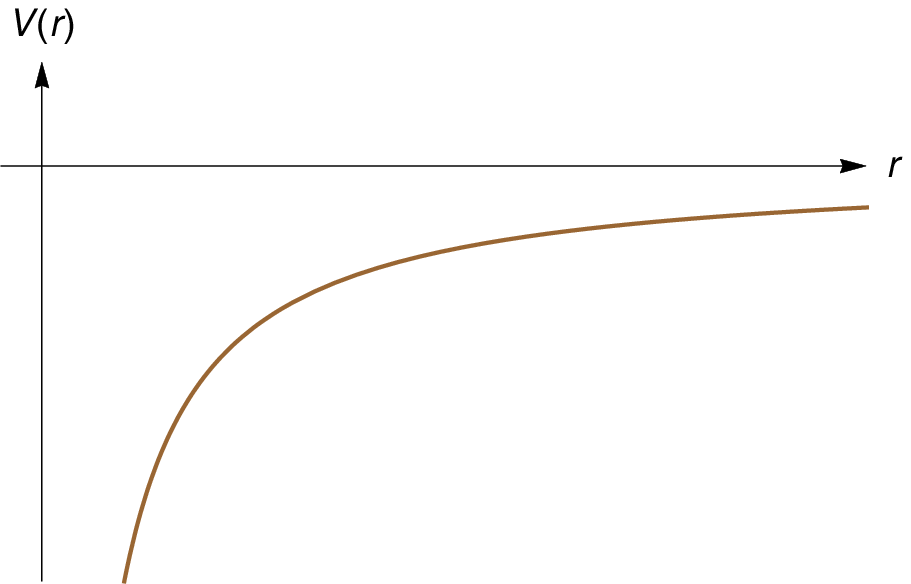}
\includegraphics[scale=0.8]{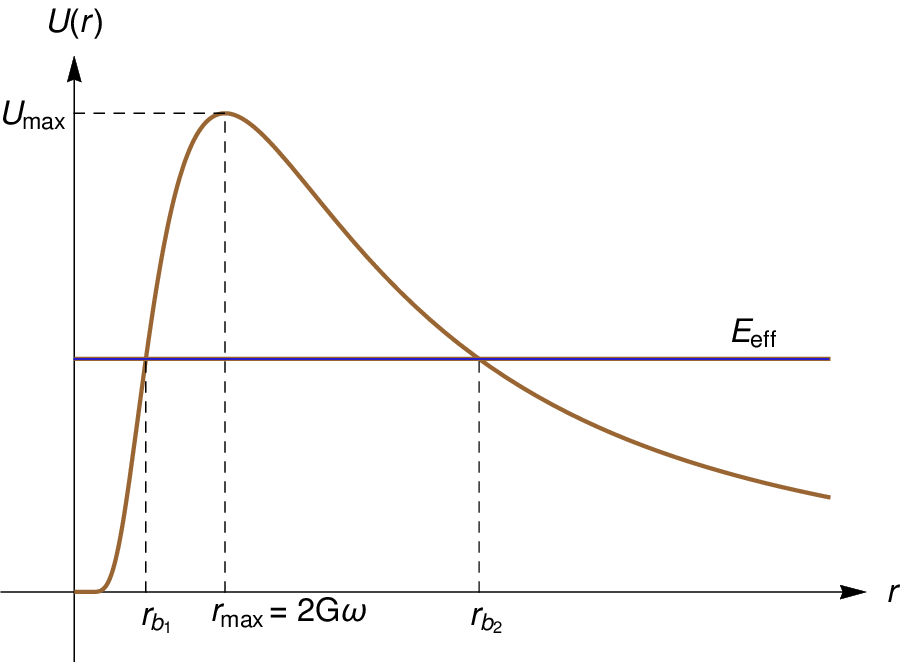}
\includegraphics[scale=0.8]{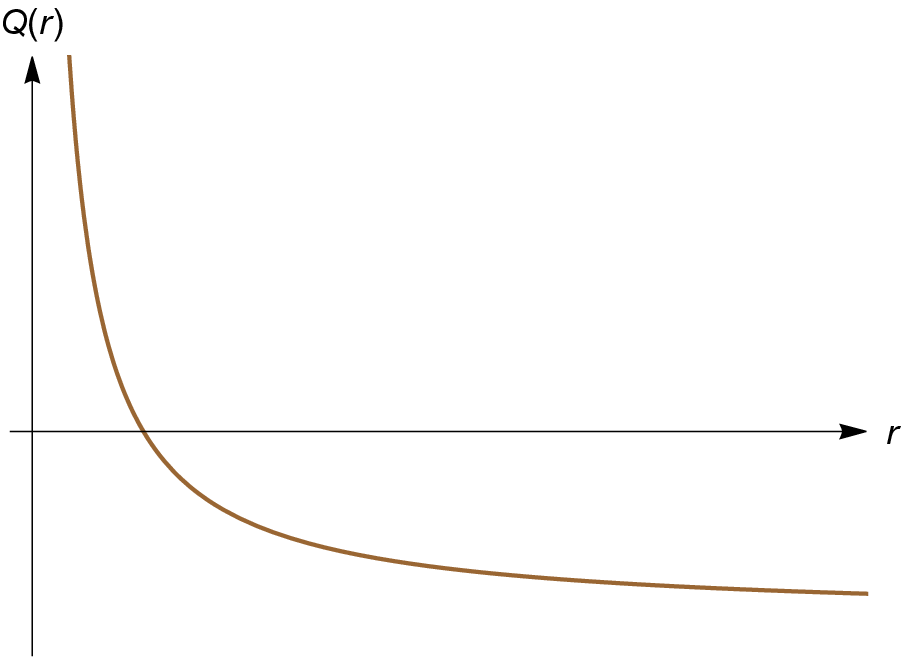}\\
\caption{Qualitative plots showing the potential $V(r)$ and the corresponding effective potential $U(r)$ for Einstein's gravity. Notice that this case is very different from the one described in Fig. 1, because here the condition (i) is violated and the potential $V(r)$ diverges at $r=0$. Therefore, the effective potential $U(r)$ goes to zero at the origin. 
It also shows the plot of the function $Q(r)$. In this case, $Q(r)$ does not possess a maximum, thus there is no local minimum for the effective potential $U(r)$ at non-zero $r$. 
If $U_{\rm max} \geqslant E_{\rm eff}$ with $r_0 < r_{b1}$ (or if $ U_{\rm max} < E_{\rm eff}$) the particles will reach $r=0$ where maybe the ``semi-classical" approximation based on Eqs.~\eq{EoM11}-\eq{EoM12} does not hold anymore. As proposed in~\cite{Guiot2020},  in this situation quantum effects may be responsible to bound the system. On the other hand, if $U_{\rm max} \geqslant E_{\rm eff}$, but $r_0 > r_{b2}$, the particles will scatter at the distance $r_{b2}$ going afterwards to the infinity. \label{EffectiveE}}
\end{figure}

Finally, before moving to the specific examples of our interest, a brief comment on Einstein's gravity is important. In the very inspiring seminal paper~\cite{Guiot2020}, the authors studied the $2 \to 2$ gravitons' scattering in Einstein's gravity by numerically solving the equation~\eq{EoM1} with interaction Newtonian potential, namely 
\be
V(r) = - \frac{4 G \om^2}{r} .
\label{V-Ein}
\ee 
The potential (\ref{V-Ein}) diverges in $r=0$ and the first assumption in~\eq{Uconditions} is violated. 
For the potential~\eq{V-Ein} the function~\eq{Q} is: 
\be
Q(r) = -1 +  \frac{2 G \om}{r} \, .
\ee
The equation $Q(r)=0$ has only one root at 
\beq
\label{rmaxEin}
r_\text{max} = 2 G \om \, , 
\eeq
where the effective potential takes the following maximum value, 
\be
\label{UmaxEin}
U_\text{max} = \frac18 \left( \frac{b}{ 2 e G \omega} \right)^2 \, e^{\frac{4G\om}{r_0}}
 \, .
\ee
Thus, for Einstein's gravity, the effective potential posses only a maximum for $r \neq 0$. In Fig.~\ref{EffectiveE}, we show the behavior of the effective potential in Einstein's gravity. For this effective potential, no bound states are allowed. The reason is that for $U_{\rm max} > E_{\rm eff}$, there is only one turning point $r_b < r_{\rm max}$, where $U(r_b) = E_{\rm eff}$. So, if $r_0 < r_{\rm max} = 2 G \om$, the only possible motion is towards $r=0$.\footnote{  The same is true for $U_{\rm max} < E_{\rm eff}$ for any value of $r_0$. For $U_{\rm max} > E_{\rm eff}$ and $r_0 > r_{\rm max}$ the particles will scatter at the turning point~$r_b' > r_{\rm max}$.}  Near $r=0$, the force grows without any bound, and the potential approximation makes no sense anymore. Considering non-linear effects, based only on a purely classical point of view, very likely the system will end up in a black hole with the particles reaching the singularity in $r=0$ in a finite amount of proper time. 
Indeed, the apparent bound states in the examples of~\cite{Guiot2020} are such that the impact parameter $b/2 \approx 2r_\text{max}$ and $U_\text{max} \lesssim E_\text{eff}$. Therefore, the system is close to the unstable equilibrium point $r_\text{max}$, and the numerical solutions of~\eq{EoM11}-\eq{EoM12} shown in~\cite{Guiot2020} will rapidly decay to $r=0$ if we allow a bigger evolution time for the system.

We conclude that according to the analytic model described in this section, there are no bound states of gravitons in Einstein's gravity as usually defined in textbooks. Indeed, the definition of bound states requires the motion to be confined between a minimum and a maximum value of the radial coordinate, namely two turning points. However, we think the analysis in \cite{Guiot2020} to be correct because if the energy is smaller than $U_{\rm max}$ and the gravitons are at a distance $r_0 < r_{\rm max}$, then they turn out to be confined. We do not know exactly how to classically interpret this bound state since the numerical integration of Eqs.~\eq{EoM11}-\eq{EoM12} fails in $r=0$ where the Newtonian potential is divergent. However, we argue that it should represent the formation of a black hole. Indeed, $r_{\rm max} = 2 G \omega$ is the Schwarzschild radius of the system. 
The same statement applies to any other theory providing a singular $V(r)$ potential in $r=0$. This situation may change due to quantum effects; see the discussion in~\cite{Guiot2020}.

Finally, we would like to stress more the main achievements of this section and the differences between the present work and Ref.~\cite{Guiot2020}. First, differently from~\cite{Guiot2020}, where the analysis of the equations of motion~\eq{EoM11}-\eq{EoM12} was done only at a numerical level, here we also obtain analytical results using the effective potential approach. The necessary and sufficient conditions for the presence of bound states are summarized in Eq.~\eq{nsconditions}. As we are going to see in the explicit examples of the next sections, the numerical solutions of Eq.~\eq{EoM11}-\eq{EoM12} are in perfect agreement with the effective potential analysis. Second, the main concern of Ref.~\cite{Guiot2020} was with a $-1/r$ potential, while here, we are interested in modified potentials $V(r)$ inspired by several classical and quantum gravity models, all of them finite at $r=0$. As discussed in the previous paragraphs, for the $-1/r$ potential, the semi-classical equations of motion may allow the particles to reach distances close to $r=0$, where full quantum effects become very important. On the other hand, for finite potentials, $V(r)$ stays bounded for any $r$. When compared with the standard $-1/r$ potential, this causes the effective potential $U(r)$ to acquire an infinite wall around $r=0$ (c.f. Figs.~1 and~2), not allowing the particles to leave the classical regime. Therefore, in opposition to Ref.~\cite{Guiot2020}, the semi-classical analysis based only on the Eqs.~\eq{EoM11}-\eq{EoM12} with the interaction potentials that we are concerned here serves as a good approximation for the description of the system in all allowed distances.

%%%%%%%%%%%%%%%%%%%%%%%%%%%%%%%%%%%%%%%%%%%%%%%%%%%%%%%%%%%%%%%%%%%
%%%%%%%%%%%%%%%%%%%%%%%%%%%%%%%%%%%%%%%%%%%%%%%%%%%%%%%%%%%%%%%%%%%

\section{A sterile scalar field coupled to nonlocal gravity: Gravi-Scalarballs}
\label{sec3}

As a first example, let us consider ordinary two-derivatives matter coupled to nonlocal gravity. 
The full action of our interest consists of a scalar field minimally coupled to nonlocal gravity, namely 
\be
\hspace{-0.3cm}
S = 
  \int  d^D x \sqrt{-g} \left[ \frac{2}{\kappa^{2}_D} \left(R + G_{\mu\nu} \gamma(\Box) R^{\mu\nu} + V(\mathcal{R})   \right)
- \frac{1}{2} g^{\mu\nu} \partial_\mu \phi \partial_\nu \phi - \frac{1}{2} m^2 \phi^2  \right] ,
\label{scalar}
\ee
where $\ka_D^2 = 32 \pi G$, $ V(\mathcal{R})$ is a ``potential'' at least cubic in the Riemann tensor and 
\be
\gamma(\Box) = \frac{e^{H(\Box)} - 1}{\Box} \, ,
\ee
where $H(z)$ is an entire function.
Since the scalar field is free in absence of gravity, namely ``sterile'', its presence does not spoil either the super-renormalizability or the finiteness of the purely gravitational theory \cite{ModestoLeslawF, ModestoLeslawR}. 

In momentum space, and ignoring the gauge dependent terms, the graviton propagator reads: 
\be
G(k) = \frac{e^{-H(-k^2) }}{i (k^2 - i \epsilon) } \left( P^{(2)} - \frac{1}{D-2} P^{(0)} \right)   ,
\label{NLP}
\ee
where $P^{(2)}$ and $P^{(0)}$ are the usual spin two and spin zero projectors operators \cite{Barnes-Rivers,Barnes-Rivers2,HigherDG}. 

The tree-level gravitational scattering amplitude for $2$-scalars into $2$-scalars can be easily obtained from the one in local gravity. For $m=0$, the result is:
\be
 A_s =  8 \pi G \frac{u t}{s} \, e^{- H(s)} \, , \qquad 
A_t =  8 \pi G \frac{s u }{t} \, e^{- H(t)} \, , \qquad %\nonumber 
 A_u=  8 \pi G \frac{s t}{u} \, e^{- H(u)} \, . 
 \label{Amplitudes}
\ee
Notice that for a proper choice of $H(z)$ the amplitudes are {\em soft} in the ultraviolet regime, namely they go to zero for large $s$, $t$ or $u$. Therefore, the unitarity bound is satisfied at tree-level. Indeed, if 
\beq
H(\Box) = (- \ell_\Lambda^2 \Box)^n \, , 
\eeq
and $n$ is an even positive integer, the tree-level amplitude turns out to be soft. On the other hand, if $n$ is an odd integer, we face with two possibilities: (1) the propagator falls off in the Euclidean, 
but the tree-level amplitude grows exponentially in the ultraviolet regime; (2) the propagator grows exponentially, but the tree-level amplitude goes to zero. Let us expand on the latter statement. For $n=1$, $H = - \ell_\Lambda^2 \Box$ and the form factor is:
\be
e^{H(\Box)} = e^{ - \ell_\Lambda^2 \Box} \quad \rightarrow \quad 
e^{H(-k^2)} = e^{ \ell_\Lambda^2 k^2}= e^{\ell_\Lambda^2 \left( - k_0^2 + \vec{k}^2 \right) } = e^{ \ell_\Lambda^2 \left( k_4^2 + \vec{k}^2 \right) } = 
e^{ \ell_\Lambda^2 k_E^2} 
 \, .
\label{n=1}
\ee
Therefore, the propagator~\eq{NLP} falls off exponentially in Euclidean space. On the other hand, the amplitudes in the three channels respectively read:
\begin{equation}
\begin{split}
& A_s =  8 \pi G \frac{u t}{s} \, e^{-  \ell_\Lambda^2 (p_1+p_2)^2 } 
=   8 \pi G \frac{u t}{s} \, e^{ \ell_\Lambda^2 s }
\, , \quad \quad
A_t =  8 \pi G \frac{s u }{t} \,
 e^{-  \ell_\Lambda^2 (p_1- p_3)^2 } 
=  - 8 \pi G \frac{s (s+ t) }{t} \, e^{ \ell_\Lambda^2 t }
\, ,  \\
&
 A_u = 
  8 \pi G \frac{s t}{u}\,
  e^{-   \ell_\Lambda^2 (p_1 - p_4)^2 } 
%  %= - \frac{g^2}{u} e^{- H(u)} 
=  8 \pi G \frac{s t}{u}\, e^{ \ell_\Lambda^2 u } =  8 \pi G \frac{s t}{u} \, e^{- \ell_\Lambda^2 (s+t) }
  \, . 
 \label{AmpliScalarNL2}
\end{split}
\end{equation}
Differently, for $n=2$ the entire function is $H = ( - \ell_\Lambda^2  \Box)^2$ and the amplitudes (\ref{AmpliScalarNL2}) turn into:
\beq
\begin{split}
& A_s =  8 \pi G \frac{u t}{s} \, e^{-  [\ell_\Lambda^2 (p_1+p_2)^2]^2 } 
=   8 \pi G \frac{u t}{s} \, e^{ - (\ell_\Lambda^2 s)^2 }
\, , 
\\
& 
A_t =  8 \pi G \frac{s u }{t}   \, 
 e^{-  [\ell_\Lambda^2 (p_1- p_3)^2]^2 } 
=  - 8 \pi G \frac{s (s+ t) }{t} \, e^{- (\ell_\Lambda^2 t )^2}
\, , \\
&
 A_u = 
  8 \pi G \frac{s t}{u}\,
  e^{-  [ \ell_\Lambda^2 (p_1 - p_4)^2]^2 } 
%  = - \frac{g^2}{u} e^{- H(u)} 
=  8 \pi G \frac{s t}{u}\, e^{ - (\ell_\Lambda^2 u)^2 } = - \frac{g^2}{s} e^{- (\ell_\Lambda^2 (s+t))^2 }
  \, ,
 \label{AmpliScalarNL3}
\end{split}
\eeq
which approach zero for $s, t \rightarrow \infty$. 

In the Regge limit $t \ll s$, which is what we need in order to compute the interaction potential~\eq{red_pot}, the leading contribution comes from the amplitude in the $t$-channel, i.e., 
\be
A_t(s,t) \approx - 8 \pi G \frac{s^2}{t} e^{- H(t)} \, . \label{matterT}\quad 
\ee
 For the amplitude (\ref{matterT}) we can now compute the potential~\eq{red_pot}. In $D=4$ we find:
 \be
&& \hspace{-1.0cm} 
V(r)
=  - \frac{8  G \om^2}{\pi r}   \int \frac{dq}{q} \, \sin (2qr)  \, 
e^{- H(-q^2) } .
\label{phaseMatter}
\ee 

As an explicit simple example we can evaluate (\ref{phaseMatter}) for the following form factor, 
\be
 e^{- \ell_\Lambda^2 \Box} \, .
\label{SFT}
\ee
As discussed above, for the string-inspired form factor (\ref{SFT}), $A_s$ and $A_t$ in (\ref{Amplitudes}) diverge for large $s$ and $t$. 
However, if we do not care about such a problem, the potential (\ref{phaseMatter}) after integration reads:
\be
V(r)= - \frac{4 G \om^2}{r} \, {\rm erf}\left( \frac{r}{\ell_\Lambda} \right)
,
\label{deltaSFT}
\ee
where ${\rm erf}(x)$ is the error function and we used $s = 4 \omega^2$. 
The potential~\eq{deltaSFT} reduces to the one in Einstein's theory when $r \to \infty$ (or equivalently $r \gg \ell_\Lambda$) 
\be
V(r) \underset{r \to \infty}{\sim} - \frac{4 G \om^2}{r} ,
\label{Vlim1}
\ee 
while at the origin, we have
\be
\lim_{r \to 0}V(r) = - \frac{8 G \omega^2}{\sqrt{\pi } \ell_\Lambda}
.
\label{Vlim2}
\ee 
Therefore, the conditions (i)--(ii) for the potential are satisfied. 

Analytical results can also be obtained for the form factors
\be
e^{ ( -\ell_\Lambda^2  \Box)^n} 
\qquad 
{\rm with} 
\qquad n>1 \, .
\label{SFT2}
\ee
For the case $n=2$ (in this case, the tree-level amplitude falls off to zero at high energy in all channels consistently with the request of having a soft amplitude in the ultraviolet regime), 
\be
%&& \hspace{-0.7cm} 
V(r)  = - \frac{8 \pi G \omega^2}{\pi \ell_\Lambda } \left[ 2 \,  \Gamma \! \left(\frac{5}{4}\right) \,_1F_3\left(\frac{1}{4};\frac{1}{2},\frac{3}{4},\frac{5}{4}; \frac{r^4}{16 \, \ell_\Lambda^4}\right) 
%\nonumber \\
%&&
 -\frac{1}{3} \frac{r^2}{\ell_\Lambda^2} \, \Gamma \left(\frac{3}{4}\right) \,_1F_3\left(\frac{3}{4};\frac{5}{4},\frac{3}{2},\frac{7}{4};\frac{r^4}{16 \, \ell_\Lambda^4 }\right) \right] \, , 
 \label{n2pot}
\ee
where ${}_p F_q$ denotes the generalized hypergeometric function. 
However, the potentials (\ref{n2pot}) and (\ref{deltaSFT}) have a very similar behavior in the ultraviolet as well as in the infrared regime. Indeed, 
\be
\lim_{r \to 0}V(r) = -\frac{16 G \omega^2 }{\pi  \ell_\Lambda} \Gamma \left(\frac{5}{4}\right),
\qquad \qquad
V(r) \underset{r \to \infty}{\sim} - \frac{4 G \om^2}{r}.
\ee
Hence, in the rest of this section, we will study the implications of the simplest potential~(\ref{deltaSFT}), being sure that the qualitative picture will not be affected by a different choice of the form factor. 
 
From the potential~\eq{deltaSFT} we get the force:
\be
F(r) = - \frac{2 G\omega^2}{r^2} \Bigl[\text{erf}(\mu r)-\frac{ 2\mu r}{\sqrt{\pi}} \, e^{-(\mu r)^2} \Bigr],
\label{N Force}
\ee
where we defined $ \mu = 1/\ell_\Lambda$ introducing the non-locality scale. 
Studying (\ref{N Force}), one can prove that:
\be
\lim_{r\to 0} F(r) = 0
\qquad
\text{and}
\qquad 
F(r) \underset{r \to \infty}{\rightarrow} -{ \frac{2 G \om^2}{r^2}}
\, .
\label{Flim1}
\ee
Thus, the force satisfies conditions (iii)-(iv). Together, the equations~\eq{Vlim1}, \eq{Vlim2} and~\eq{Flim1} imply that~\eq{Uconditions} and \eq{Qconditions} hold true. So, the presence of bounded solutions is related to the existence of the positive maximum for the function~\eq{Q}, $Q_\text{max} > 0$.

Using~\eq{N Force} we get for~\eq{Q}
\be
Q(r) = - 1 + 2 \, G \mu \, \om S(\mu r) \, ,
\label{Q-nonlocal-1}
\ee
where the function $S(x)$ is defined by:
\be
S(x) = \frac{\text{erf}(x)}{x}
- \frac{2}{\sqrt{\pi}} \, e^{-x^2}
\label{V}
\, .
\ee
\begin{figure}[t]
\includegraphics[scale=0.6]{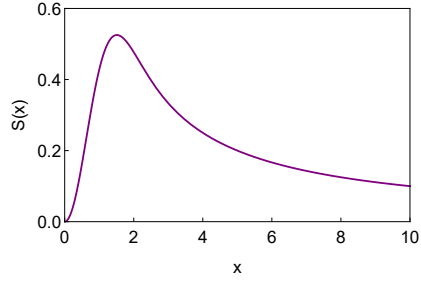}
\caption{\small Plot of the function~\eq{V}.\label{alphaVS}}
\end{figure}
\!\!The behavior of~\eq{V} is shown in Fig.~\ref{alphaVS}. It has a maximum value $S_\text{max} = 0.53$ at the point $x_\text{max} = 1.51$. Therefore, the first condition in~\eq{nsconditions} is:
\be
Q_\text{max} = - 1 + 2 \, G \mu \om S_\text{max} > 0
.
\ee
By solving the above inequality, we can find the following bound on the energy, 
\be
\om  > \frac{1}{2 S_\text{max} \, G \mu }
\label{w-grav-nonlocal}
\, .
\ee
Let us assume that the non-locality scale is of the order of the Planck mass, i.e. $\mu \sim M_\text{P}$. Then, using $G = 1/M_\text{P}^2$ Eq.~\eq{w-grav-nonlocal} implies that:
\be
\om  > 0.95 \, M_\text{P}
\label{w-grav-nonlocal-2}
.
\ee
If the condition~\eq{w-grav-nonlocal} is satisfied, the equation $Q(r) = 0$ has two solutions and, as discussed in Section~\ref{sec2}, the effective potential~\eq{effective potential energy} has a local minimum at the point $r_\text{min}$ and a local maximum at $r_\text{max} > r_\text{min}$. 
We can not find general analytic expressions for $r_\text{min}$ and $r_\text{max}$, 
but it is possible to find numerically values for them for fixed value of the parameters $\mu$ and $\om$. However, using~\eq{d2W}, it is possible to find some constraints on $r_\text{min}$ and $r_\text{max}$ by studying the first derivative of $Q$ at $r = r_e$
\be
\frac{dQ}{dr} \Big|_{r = r_e} = \frac{2 G \omega}{r_e^2} \, T(\mu \, r_e) \, ,
\ee
where we introduced the function: 
\be
T(x) = \frac{2x(2x^2+1)}{\sqrt{\pi}} \,e^{-x^2}-\text{erf}(x) \, .
\label{X beta}
\ee
\begin{figure}[t]
\includegraphics[scale=0.6]{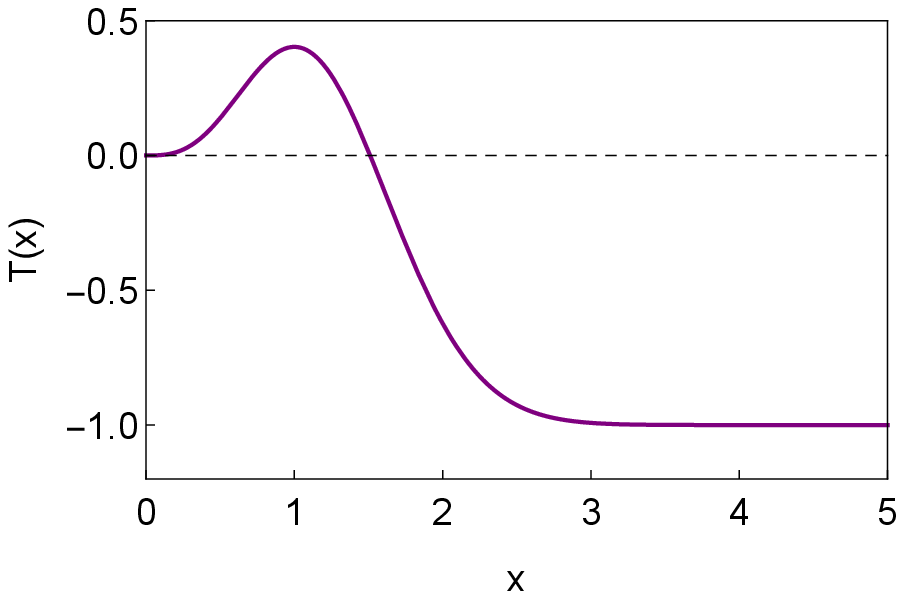}
\caption{\small Plot of the function~\eq{X beta}.\label{betaX}}
\end{figure}
\!\!\!\!
The plot of $T(x)$ is shown in Fig.~\ref{betaX}. Solving numerically the equation $T(x)=0$, we find that the root is approximately at $x = 1.51$. Therefore, the function $T(\mu r_e)$ is positive for $0 < r_e  <1.51/\mu$ and negative if $r_e > 1.51/\mu$. Thus, according to~\eq{d2W2} we have:  
\be
0 <r_\text{min} < \frac{1.51}{\mu}< r_\text{max} \, .
\label{rmin rmax gs}
\ee
The above equation gives a lower and an upper bound for $r_\text{min}$, but only a lower bound for $r_\text{max}$. To found an upper bound on $r_\text{max}$ we study how the function $Q(r)$ depends on the non-locality scale $\mu$. Taking the partial derivative with respect to $\mu$ we get:
\be
\frac{\pa Q}{\pa \mu} = \frac{8 G \omega \mu ^2 r^2}{\sqrt{\pi }} \, e^{-\mu ^2 r^2 } >0
\, .
\ee
Namely, $Q$ is a monotonously increasing function of $\mu$ for any fixed $r$. Therefore, the maximum values for $Q$ are obtained in the limit $\mu \to \infty$, when the potential~\eq{deltaSFT} reduces to the one of the Einstein's theory~\eq{V-Ein}. Hence, the possible values for $r_\text{max}$ for the non-local theory are bounded by the maximum radius of the local theory, which is given in \eq{rmaxEin}.

\begin{figure}[t]
(A-1)\hspace{4cm}
(B-1)\hspace{4cm} (C-1)\\
\includegraphics[width=4.2cm]{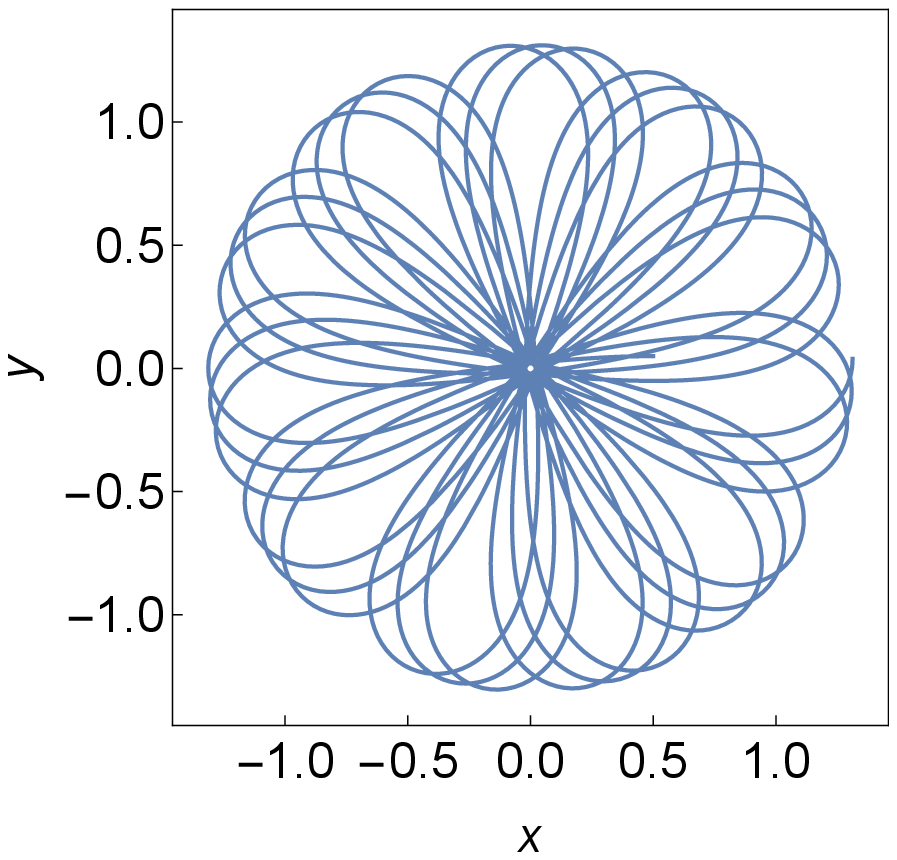}
\hspace{0.5cm}
\includegraphics[width=4.2cm]{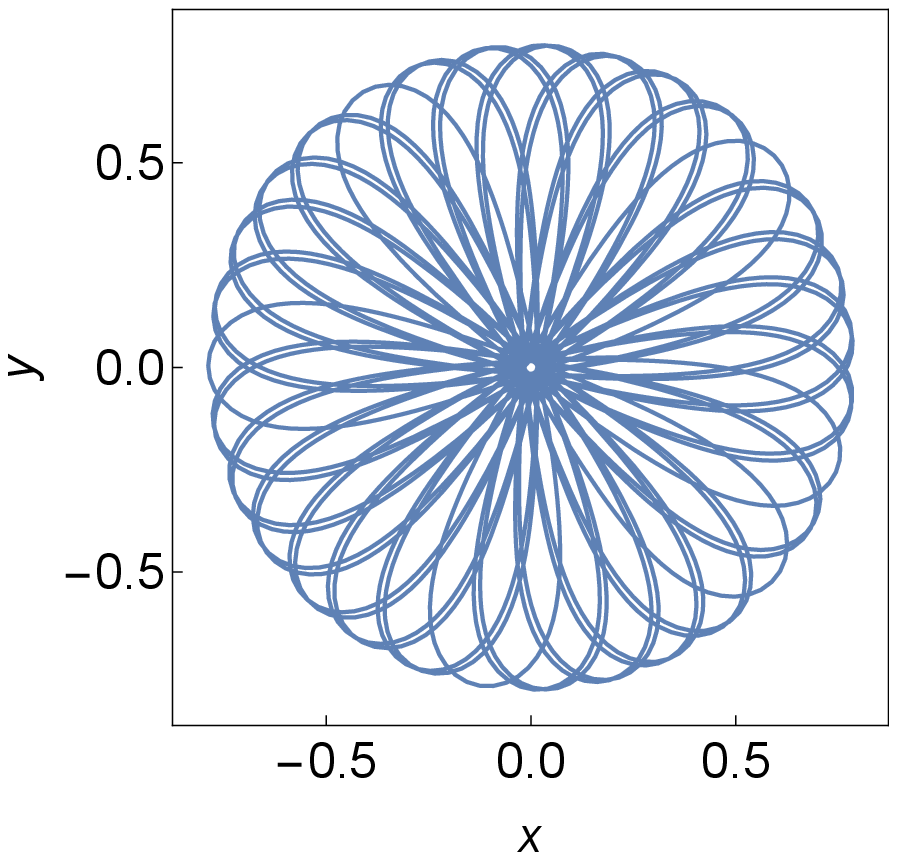}
\hspace{0.5cm} 
\includegraphics[width=4.15cm]{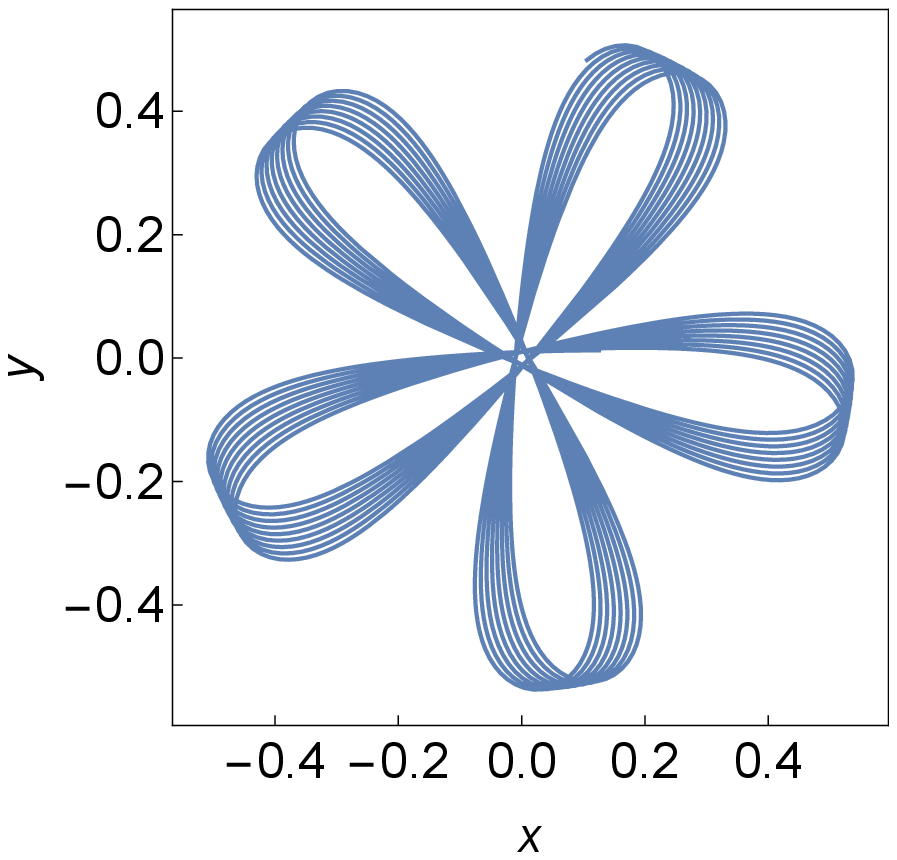}
\\
(A-2)\hspace{4cm} (B-2)\hspace{4cm} (C-2)\\
\includegraphics[width=4.2cm]{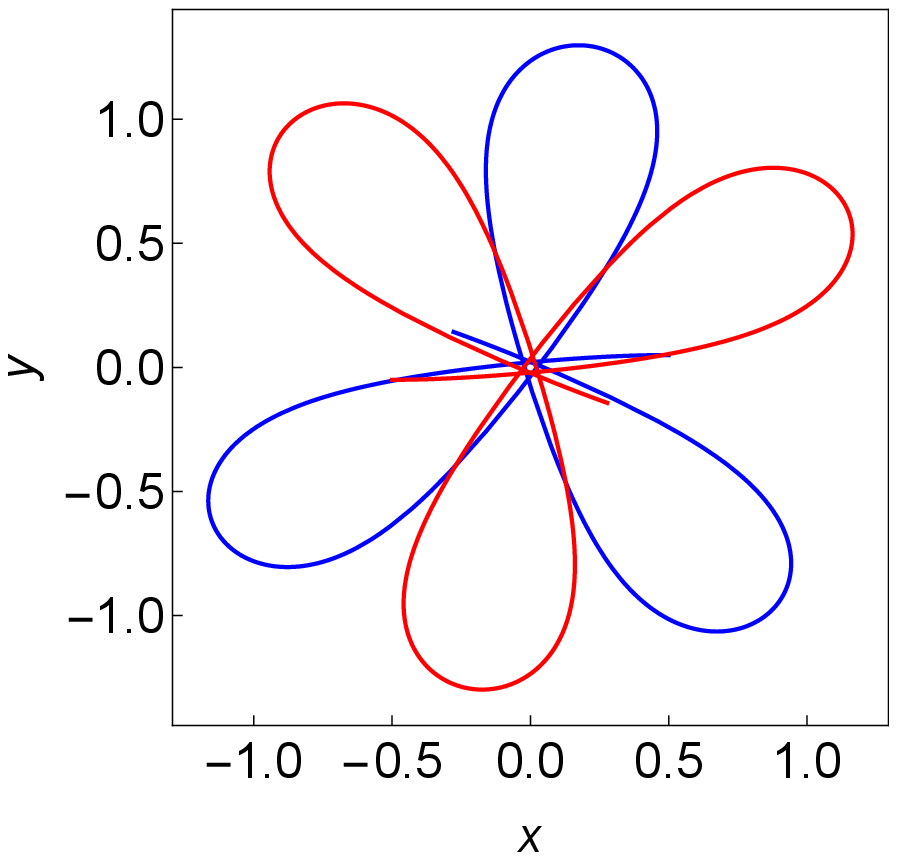}
\hspace{0.5cm}
\includegraphics[width=4.2cm]{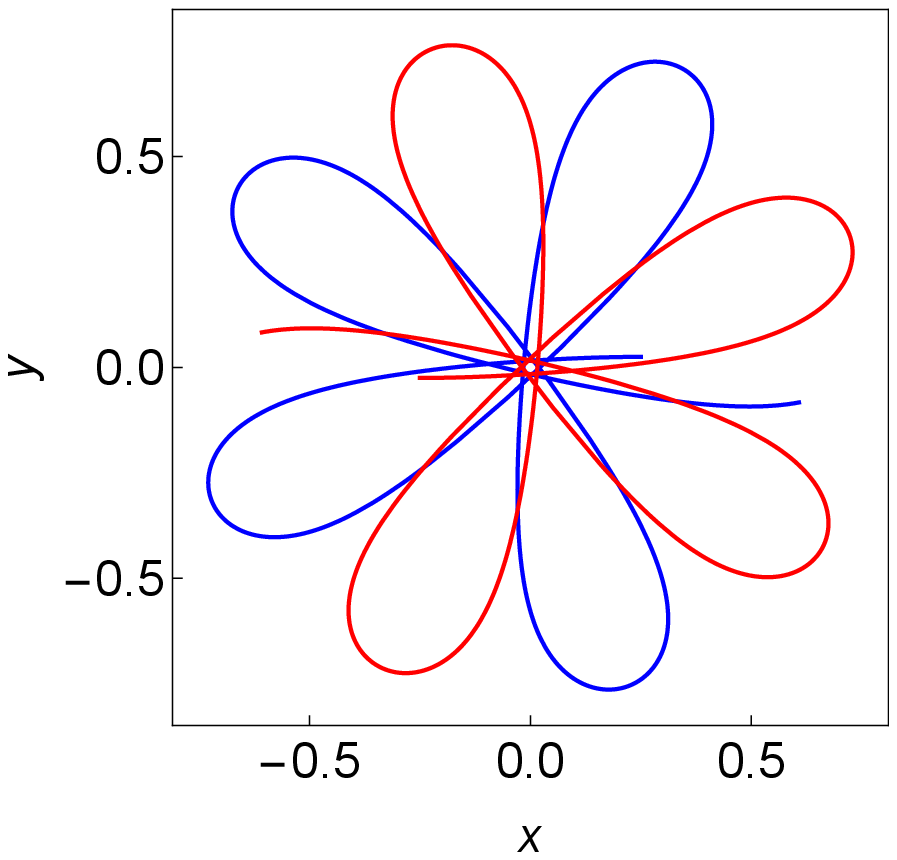}
\hspace{0.5cm} 
\includegraphics[width=4.2cm]{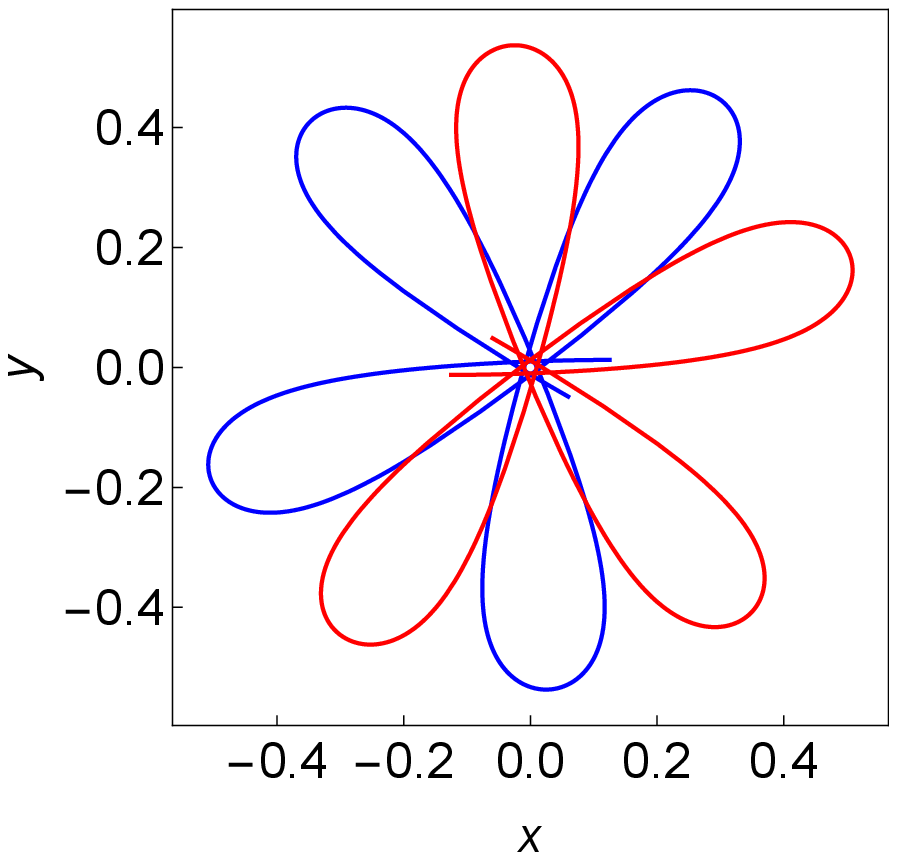}
\\
\caption{Three examples of bound states for $\mu = M_\text{P} = 1$. 
(A-1) shows the trajectory in the reduced coordinate $\vec{r}$ (\ref{reducedR}) for $t_\text{f}=100$, and (A-2) shows the trajectories for the two particles with $t_\text{f}=10$. 
In both cases $\omega=5$,  $a=1$, and $b=0.1$. 
(B-1) shows the trajectory in the reduced coordinate $\vec{r}$ for $t_\text{f}=100$, and (B-2) % 
for $t_\text{f}=8$. Here $\omega=10$, $a=0.5$, and $b=0.05$. (C-1) Trajectory in the reduced coordinate $\vec{r}$  for $t_\text{f}=50$, and (C-2) for $t_\text{f}=5$. Here $\omega=20$,  $a=0.25$, and $b=0.025$.  
In blue, we displayed the trajectory of the particle of radial vector $\vec{r}_1$, while in red, we displayed the particle with radial vector $\vec{r}_2=-\vec{r}_1$.
}
\label{NTrajectory}
\end{figure}

Therefore, the necessary and sufficient conditions for the existence of bound states are: 
\beq
\begin{split}
&
\boxed{\om >  \frac{0.95}{G \mu},
\qquad
U_\text{max} > \frac12,
\qquad
r_{b_1} < r_0 < r_{b_2}} \, ,
\\
&
\boxed{
0 < r_{b_1} <r_\text{min} < r_{b_2}  < r_\text{max},
\qquad
r_\text{min} < \frac{1.51}{\mu} < r_\text{max}<2 G\omega}
\, . 
\label{constraint gs}
\end{split}
\eeq
In Fig.~\ref{NTrajectory} we show some examples of bound states by numerically integrating the equations \eqref{dotR} and \eqref{dotphi} from $t_0=0$ until a finite time $t_\text{f}$ and plotting the particles' trajectory in $xy$-plane with the aid of~\eqref{polar}.

{\em We conclude that in nonlocal gravity, the necessary condition for having bound states of sterile scalars is $\omega \gtrsim M_{\rm P}$, and they have an extension $r_{b_2}$ that decreases with increasing of the energy for fixed $r_0$ and $b$} . Moreover, the initial conditions must satisfy $b \ll a$ and $r_{b_1} < r_0 < r_{b_2}$ consistently with $t \ll s$.
{\em Since the existence of bound states is based on a scalar field coupled to nonlocal gravity, we call these bound states: gravi-scalarballs. }

A very similar result can be obtained in general \cite{shapiro3}, or Lee-Wick \cite{ShapiroModestoLW, ModestoLW} higher derivative gravitational theories. The computation consists of merging together the results of Sec. \ref{sec-scallar} and the theory (\ref{scalar}) with the form factor $\ga(\Box)$ replaced by a polynomial.

In this section, the interaction potential has been defined by means of tree-level scattering amplitudes. Therefore, one may wonder if loop corrections can spoil our result and bound states do not form. To answer this question, we need to understand if the good properties (i)-(ii) for the potential, described in Sec.~\ref{sec2}, are destroyed by the loop contributions. It turns out that the answer is actually negative. Indeed, typically the one-loop quantum corrections have a Universal form in the ultraviolet regime; namely, they lead to include a $k^4 \log(k^2/\mu^2)$ term in the denominator of the propagator. However, such correction can only improve the convergence of the tree-level potential $V(r)$ at $r=0$~\cite{Burzilla:2020utr,Burzilla:2020bkx}. In the far-infrared regime, the one-loop quantum corrections to the potential have also an Universal form leading to the correction $1/r^3$ to the potential~\cite{Burzilla:2020utr,Burzilla:2020bkx,Donoghue:1994dn,Helayel-Neto:1999ryv,dePaulaNetto:2021axj}. However, at large distances $1/r^3$ is sub-leading with respect to the classical counterpart $1/r$. We expect that higher-loops contributions will also not spoil the conditions (i)-(ii) because of the asymptotic freedom \cite{Briscese:2019twl, Rachwal:2021bgb} in super-renormalizable theories \cite{Krasnikov, Modesto, ModestoLeslawR} or the softness of the quantum amplitudes in finite theories \cite{Kuzmin, ModestoLeslawF, Modesto:2021okr, Modesto:2021ief,MeSpacetime-Matter}. 

Therefore, loop corrections will not change the main qualitative conclusions on bound states here described.

%%%%%%%%%%%%%%%%%%%%%%%%%%%%%%%%%%%%%%%%%%%%%%%%%%%%%%%%%%%%%%%%%%%
%%%%%%%%%%%%%%%%%%%%%%%%%%%%%%%%%%%%%%%%%%%%%%%%%%%%%%%%%%%%%%%%%%%

\section{Scattering Amplitudes in String Theory: Stringballs}
\label{sec-string}

One of the main properties of string theory is the softness of the tree-level scattering amplitude at high energy~\cite{Polchinskibook}. 
Therefore, string theory naturally has the right feature needs in our scenario of dark matter as bound states. 

Let us start reminding the following very compact and suggestive Veneziano-tree-level scattering amplitude in closed string theory (see \cite{Camanho:2014apa} and references in within), 
\be
A^{\rm (string)} = \frac{\Gamma\left( - \frac{\alpha^{\prime}}{4} s \right) \Gamma\left( - \frac{\alpha^{\prime}}{4} t \right) \Gamma\left( - \frac{\alpha^{\prime}}{4} u \right)}{
\Gamma\left( 1 +  \frac{\alpha^{\prime}}{4} s \right)
\Gamma\left( 1 + \frac{\alpha^{\prime}}{4} t \right)
\Gamma\left( 1 + \frac{\alpha^{\prime}}{4} u \right)
}
%\, 
%A^{\rm (field \, theory)}
\, ,
\label{CSA}
\ee
(a similar result appears for the open string theory amplitudes). We now focus on the string % 
amplitude (\ref{CSA}) and we take the limits: $t \ll s$, $s \gg 4/\alpha^\prime$, $t \ll 4/\alpha^\prime$ \cite{Camanho:2014apa, Siegel:2003vt} (see also \cite{Amati:1987wq, Amati:1987uf, Amati:1988tn, Amati:1992zb, Amati:1993tb}). In the latter approximations the amplitude (\ref{CSA}) reads:
\be
A^{\rm (string)} \approx  \frac{\Gamma\left( - \frac{\alpha^{\prime}}{4} t \right)}{\Gamma\left( 1 + \frac{\alpha^{\prime}}{4} t \right)
}
\, \left( -  {\rm i} \frac{\alpha^\prime}{4} s \right)^{-2 + \frac{\alpha^\prime}{2} t} 
%A^{\rm (field \, theory)} 
= -
 \frac{\Gamma\left( - \frac{\alpha^{\prime}}{4} t \right)}{\Gamma\left( 1 + \frac{\alpha^{\prime}}{4} t \right)
}
\, \left( \frac{\alpha^\prime}{4} s \right)^{-2 + \frac{\alpha^\prime}{2} t} \, ( - {\rm i } )^{\frac{\alpha^\prime}{2} t} 
%\, %A^{\rm (field \, theory)}
\, .
\label{CSAapprox}
\ee
In the last expression we properly extracted the factor $( - {\rm i } )^{\frac{\alpha^\prime}{2} t}$ from which we get an imaginary contribution to the amplitude. Such contribution is saying that the most likely process is to create a massive closed string rather than the scattering of particles. Therefore, in order to avoid such an inelastic regime, we have to assume again small $t$ and the final amplitude further simplifies to:
\be
A^{\rm (string)} \approx    
  \frac{4}{\alpha^{\prime}  t}  \, \left( \frac{\alpha^\prime}{4} s \right)^{-2 + \frac{\alpha^\prime}{2} t} \,  
%A^{\rm (field \, theory)} 
=
\frac{64}{\alpha^{\prime 3}  s^2 t}  \, \left( \frac{\alpha^\prime}{4} s \right)^{ \frac{\alpha^\prime}{2} t} \,  
%A^{\rm (field \, theory)} 
\, . 
\label{CSAapprox2}
\ee
Now we can compute the potential for the scattering of gravitons in string theory after selecting a suitable front coefficient in the proper generalization of formula (\ref{phaseMatter}) in string theory,
\be
 %\hspace{-1.0cm} 
V(r) & = & \# \int dq \, \frac{\sin(2 q \, r)}{q\, r} \, A^{\rm (string)}(s, - q^2 ) 
=
 \#  \int dq \, 
  \frac{\sin(2 q \, r)}{q\, r}  
   \, 
 \left( - \frac{64}{\alpha^{\prime 3}  s^2 q^2} \right)  \, \left( \frac{\alpha^\prime}{4} s \right)^{ - \frac{\alpha^\prime}{2} q^2}
  \nonumber\\
 & = & 
 - \frac{G s}{r} \, {\rm erf }\left[ \frac{r}{  \sqrt{ \frac{\alpha^\prime}{2} \ln \left( \frac{ \alpha^\prime s}{4}  \right)  } }\, \right]
 \quad {\rm for} \quad s \gg \frac{4}{\alpha^\prime} 
 \, ,
\label{stringPote}
 \ee
where $\# = G \alpha^{\prime 3} s^3/32 \pi$ is the right coefficient according to the infrared Newtonian regime. Taking  $s = 4\omega^2$, we can rewrite Eq.~\eqref{stringPote} as:
\be
V (r) = - \frac{4G\omega^2}{r} \, \text{erf}\left[M(\omega,\alpha') r \right]
\, ,
\label{Vstring2}
\ee
where according to $s \gg 4/\alpha'$ (\ref{stringPote}), 
\be
\omega^2 \gg \frac{1}{\alpha'} \, ,
\label{oma}
\ee
and we defined the energy-dependent mass scale:
\be
M(\omega,\alpha')=\sqrt{\frac{2}{\alpha'\ln(\alpha'\omega^2)}} \, .
\label{M omega alpha} 
\ee
Comparing Eq.~\eqref{Vstring2} with \eqref{deltaSFT}, we see that the only difference between them is the % 
presence of the energy $\omega$ 
inside the error function. Therefore, all the results of Section~\ref{sec3} remain true under the replacement of the parameter $\mu = 1/\ell_\La$ with $M(\omega,\alpha')$. Thus, the bound on the energy $\om$ can be obtained through~\eq{w-grav-nonlocal} with the replacement $\mu \mapsto M(\omega,\alpha') $
\bea
\omega > \frac{1}{2S_\text{max} G M(\omega,\alpha')} \, , \qquad S_\text{max} = 0.53 \, .
\label{NCS}
\eea
Unlike in Sec. \ref{sec3} here $M$ is a function of the energy $\om$. Hence, we need to study with care Eq. \eq{NCS} in order to obtain the bound on $\om$. Replacing~\eqref{M omega alpha} into~\eqref{NCS} we get the following inequality: 
\be
h(\om) < 0 \, ,
\quad \text{where} \quad 
%\label{NCS2}
%\ee 
%\be
h(\omega)=\frac{\ln(\alpha'\omega^2)}{\alpha'\omega^2}
-\frac{8 G^2 S_\text{max}^2}{\alpha'^2} \, .
\label{NCS2_h}
\ee
The function $h(\omega)$ in~\eq{NCS2_h} has the asymptotic limits, 
\be
\label{hass}
\lim_{\om \to 0} h(\om) = - \infty \, ,
\qquad 
\lim_{\om \to \infty} h(\om) = - \frac{8 G^2 S_\text{max}^2}{\alpha'^2} \, , 
\ee
and the equation $h'(\omega) = 0$ has only one solution for $\omega$ equals to:
\be
\omega_\text{max} = \sqrt{\frac{e}{\alpha'}} \, ,
\label{wmax}
\ee 
where the function $h(\om)$ takes the maximum value
\be
h_\text{max} = \frac{1}{e} - \frac{8 G^2 S_\text{max}^2}{\alpha'^2} \, .
\ee 
In the above equations $e = 2.718 \dots$ is the Euler number. Therefore, it is possible to split the study of the inequality in \eqref{NCS2_h} into two distinct cases. 
\begin{itemize}
\item[(I)]
If $\alpha' \leqslant 2\sqrt{2 e }G \, S_\text{max}$, then $h_\text{max} \leqslant 0$ and the inequality~\eqref{NCS2_h} holds for all values of $\omega$ because of~\eq{hass}. In this case, the only bound on $\om$ is given by~\eq{oma}. Therefore, the necessary conditions for bound states are:
\be
\boxed{ \alpha^\prime \leqslant 2.47 G \, ,
\qquad 
\omega \gg \frac{1}{\sqrt{\alpha'}} } \, .
\label{string-1}
\ee
\item[(II)] 
If $\alpha' > 2\sqrt{2 e }G \, S_\text{max}$, then $h_\text{max} > 0$ and the equation $h(\omega) = 0$ has two distinct solutions. We denote the roots of $h(\omega) = 0$ as $\omega_1(\alpha')$ and $\omega_2(\alpha')$ with $\omega_2(\alpha') > \omega_1(\alpha')$. Clearly, $\omega_1(\alpha')< \omega_\text{max} <\omega_2(\alpha')$, then using Eq.~\eq{wmax} we get:
\be
\label{waw}
\omega_1(\alpha')<\sqrt{ \frac{e}{\alpha'} }<\omega_2(\alpha') \, 
.
\ee
In the region $\omega_1(\alpha')<\omega<\omega_2(\alpha')$, the function $h(\omega)$ is always positive. On the other hand, for $0<\omega<\omega_1(\alpha')$ and $\omega>\omega_2(\alpha')$ the function $h(\omega)$ is negative. Considering the inequality~\eqref{oma}, we can safely take $ \omega> \sqrt{e/\alpha'} $ and given~\eq{waw} the condition $h(\omega)<0$ is reduced to $\omega > \omega_2(\alpha')$. Accordingly, the necessary conditions for having bound states are:
\be
\boxed{\alpha^\prime > 2.47 G \, ,
\qquad \omega \gg \frac{1}{\sqrt{\alpha'}}, 
\qquad 
\omega > \omega_2(\alpha')
} \, 
. 
\label{string-2}
\ee
\end{itemize}
The bounds on the maximum and minimum radius are the same as in Section~\ref{sec3} with the replacement $\mu \mapsto M(\omega,\alpha') $. 

Hence, in conclusion, the necessary and sufficient conditions for having bound state solutions are~\eq{string-1} or~\eq{string-2} together with    
\beq
\begin{split}
\boxed{U_\text{max} > \frac12 \, ,
\qquad
r_{b_1} < r_0 < r_{b_2} \, ,
\qquad
0 < r_{b_1} < r_\text{min} < r_{b_2} < r_\text{max} \, ,
\qquad
r_\text{min} < \frac{1.51}{M(\omega,\alpha')} < r_\text{max}< 2 G\omega }  \, .
\label{string_con}
\end{split}
\eeq
\begin{figure}[t]
(a)\hspace{4cm} (b)\\
\includegraphics[width=4cm]{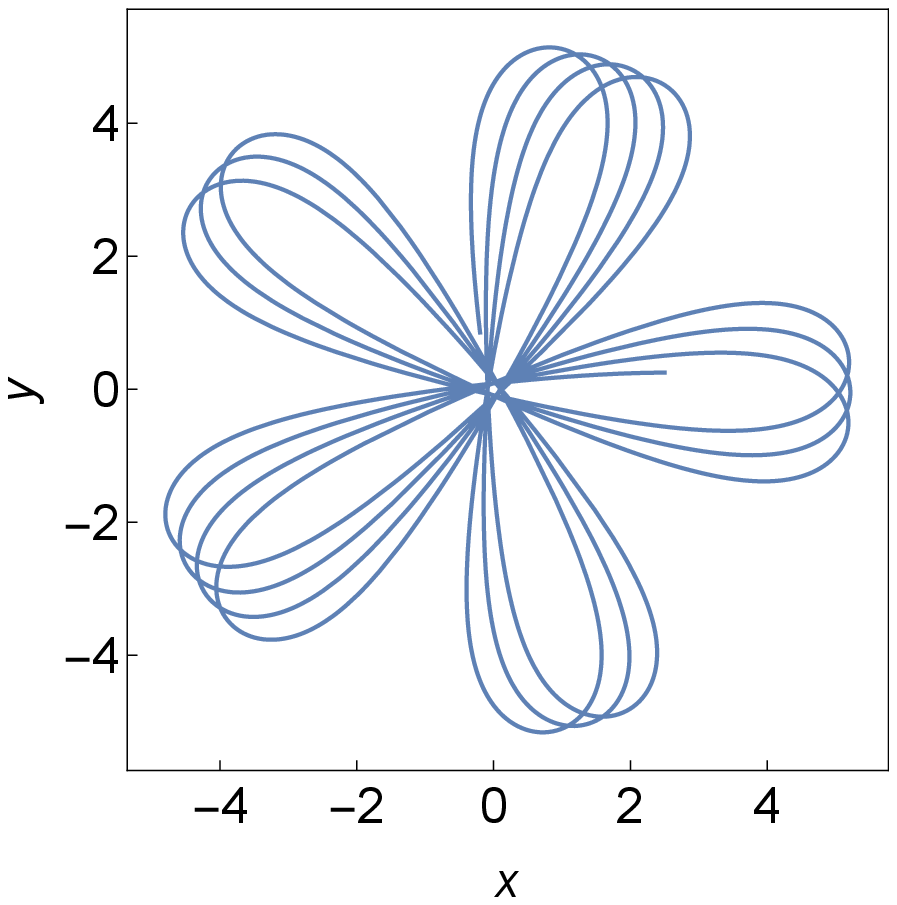}
\hspace{1cm}
\includegraphics[width=4cm]{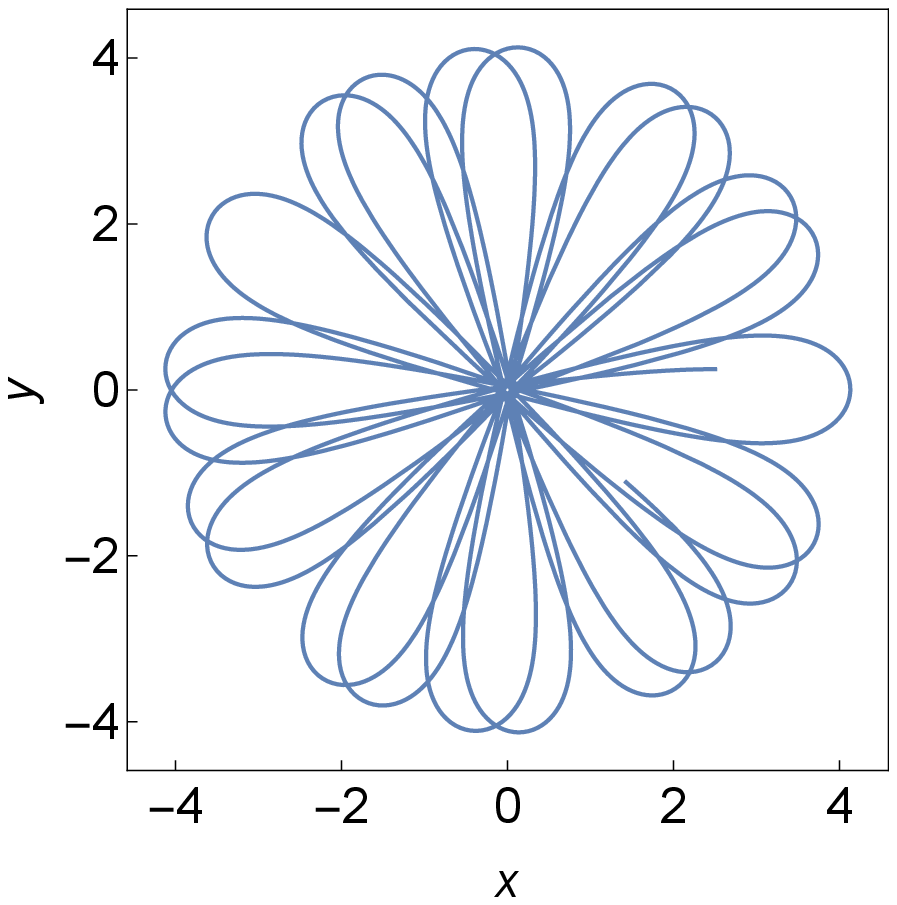}
\caption{\small We here show two examples of bound states for $\alpha'=10$: (a) $\omega=100$,  $a=5$, $b=0.5$ and $t_\text{f}=200$; (b) $\omega=200$,  $a=5$, $b=0.5$ and $t_\text{f}=200$. 
\label{StringTrajectory}
}
\end{figure}
\begin{table}[t] 
(a)\hspace{5cm} (b)\\
\begin{tabular}{|l|c|c|c|c|}
\hline    
    & 
    $\, M(\omega,\alpha') \,$
    &
    $\,1.51/M(\omega,\alpha')\,$
    &
    $\, r_\text{min} \,$
    &
    $\, r_\text{max} \,$
    \\
    \hline
    %2
    $ \omega = 100 $
    & $ 0.13 $
    & $ 11.46 $
    & $ 1.73 $
    & $ 199.99 $
    \\
    \hline
    %1
    $ \omega = 200 $
    & $ 0.12 $
    & $ 12.13 $
    & $ 1.32 $
    & $ 399.99 $
    \\
    \hline
\end{tabular}
%\caption{}
%\label{Table-string-a}
%\end{subtable}
\hspace{1cm}
%\begin{subtable}{.4\textwidth}
%\centering
\begin{tabular}{|l|c|c|c|c|c|c|}
\hline    
    & 
    $\,a \,$
    &
    $\,b\,$
    &
    $\,r_0\,$
    &
    $\,r_{b_1}\,$
    &
    $\,r_{b_2}\,$
    &
    $U_\text{max}$
    \\
    \hline
    %1
    $ \omega = 100 $
    & $ 5 $
    & $ 0.5 $
    & $ 2.51 $
    & $ 8.74 \times 10^{-2} $
    & $ 5.22 $
    & $ 8.83\times 10^{17} $
    \\
    \hline
    %2
    $ \omega = 200 $
    & $ 5 $
    & $ 0.5 $
    & $ 2.51 $
    & $ 4.21 \times10^{-2} $
    & $ 4.13 $
    & $ 4.91\times10^{39} $
    \\
    \hline
\end{tabular}  
\caption{\small  Examples of parameters that satisfy the condition for having bound states.}
\label{Table-string-2}
\end{table}

Finally, we consider an explicit example. We choose to work in Planck units, i.e. $1/\sqrt{G} = M_\text{P} =1$ and we consider $\alpha'=10$ so that we are in the situation described in (II). Numerically solving the equation $h(\om) = 0$ we find the value for the highest root $\omega_2(\alpha') =  4.95$. Thus, the condition~\eq{string-2} is satisfied, for example, for $\omega=100$ or $\omega = 200$. For this values of the energy $\omega$ we obtain the values for $r_\text{min}$ and $r_\text{max}$ shown in Table~\ref{Table-string-2} (a). As we can see the 
condition $r_\text{min} < {1.51}/{M(\omega,\alpha')} < r_\text{max}<2 \omega$ holds. In the examples of Table~\ref{Table-string-2} (a) we found that $r_\text{max} \approx 2\omega$ because the values of $\omega$ are large when compared to $M(\omega,\alpha')$. Moreover, for a given choice of initial conditions, all inequalities in~\eq{string_con} can be satisfied. See, for example, Table~\ref{Table-string-2} (b).
In Fig.~\ref{StringTrajectory} we plot the trajectory in the $xy$-plane for the initial values described in the Table~\ref{Table-string-2} (b).

{\em We conclude that in string theory bound states of gravitons form for any $\omega \gg 1/\sqrt{\alpha'}$ if the magnitude of $\sqrt{\alpha'}$ is comparable with the Planck length\footnote{If the magnitude of $\sqrt{\alpha'}$ is larger than Planck length, this conclusion may change. Indeed, in the case (II) $\omega>\omega_2(\alpha')$ is required and we need to compare $\omega_2(\alpha')$ with $1/\sqrt{\alpha'}$. 
It turns out that 
$\omega \gg 1/\sqrt{\alpha'}$ implies $\omega>\omega_2(\alpha')$ only if $\omega_2(\alpha')$ and $1/\sqrt{\alpha'}$
are of the same order of magnitude.
E.g., for $\alpha'=100$ and $G=1$, $1/\sqrt{\alpha'}=0.1$  
and we found numerically $\omega_2(\alpha')=21.9$, such that $1/\sqrt{\alpha'}\ll\omega_2(\alpha')$. On the other hand, if the value of $\sqrt{\alpha'}$ is close to the Planck length, for example, $\alpha'=3$, the conclusion in the text remains valid because $1/\sqrt{\alpha'}=0.6$, which is comparable with 
$\omega_2(\alpha')=0.98$. 
}, 
and they have extension $r_{b_2}$ that decreases with increasing of the energy for fixed $r_0$ and impact parameter $b$}.  Moreover, the initial conditions must satisfy $b \ll a$ and $r_{b_1} < r_0 < r_{b_2}$ consistently with $t \ll s$.

{\em Since the result is based on string theory, let us call these bound states: stringballs. }

%%%%%%%%%%%%%%%%%%%%%%%%%%%%%%%%%%%%%%%%%%%%%%%%%%%%%%%%%%%%%%%%%%%
%%%%%%%%%%%%%%%%%%%%%%%%%%%%%%%%%%%%%%%%%%%%%%%%%%%%%%%%%%%%%%%%%%% 

\section{Nonlocal scalar electrodynamic: Electroballs}
\label{Electroballs}

We here investigate the feasibility of bound states in nonlocal electrodynamics coupled to a complex scalar. 
This example will turn out to be useful in Section (\ref{Planckballs}) and in the conclusions where we will generalize our results to all the fundamental interactions described by a local or nonlocal ultraviolet completion of the standard model of particle physics. In particular, we here discuss the case of a dimensionless coupling constant as a toy model for the nonlocal standard model. 

The action for the case of a massless complex scalar reads:
\be
\mathcal{L} =  - \frac{1}{4 e^2} F_{\mu\nu} e^{H( \Box_\Lambda)} F^{\mu\nu} 
- (D_\mu \Phi)^{\dagger} (D^\mu \Phi)\, , 
%\quad 
%\sum_{a}^{N_f}\bar \psi_a \, i \slashed{\cal D}  e^{H(-\slashed{\cal D}^{2}_{\Lambda})}\, \psi_a
\label{NLED}
\ee
where $D_\mu = \partial_\mu + i e A_{\mu}$, and $\Lambda =1/\ell_\Lambda$ is the non-locality scale. We do not need to introduce the covariant d'Alembertian in the form factor for the electromagnetism action (\ref{NLED}) because $F_{\mu\nu} = \partial_\mu A_\nu - \partial_\nu A_\mu$ is gauge invariant. 
The theory is the gauge analog of the one presented in section (\ref{sec3}). Indeed (\ref{NLED}) is super-renormalizable with only one-loop divergences proportional to $F_{\mu\nu} F^{\mu\nu}$. In the jargon of section (\ref{sec3}), the scalar $\Phi$ is sterile because it is non-interacting in the absence of the gauge field. 

Following again the seminal paper \cite{Guiot2020} we focus on the 
$\pi^+ \, \pi^- \rightarrow \pi^+ \, \pi^-$ interaction, whose scattering amplitude in the Regge's limit $t \ll s$ reads: 
\begin{equation}
%\mathcal{M}^{\text{tree}}=
A_{t} \approx - 2 i  e^2 \frac{s}{t} e^{-H(t)} \,  .
\end{equation} 
Comparing the above amplitude with (\ref{matterT}), we get the potential (\ref{deltaSFT}) with in it the replacement $8 \pi  G s = 16 \pi  G \omega^2 \rightarrow e^2$. The potential reads:
%(\ref{red_pot})
\begin{equation}
V(r)= - \frac{4 G \om^2}{r} \, {\rm erf}\left( \frac{r}{\ell_\Lambda} \right) \quad \rightarrow \quad 
V(r) = - \frac{e^2}{4\pi r} \, {\rm erf}\left( \frac{r}{\ell_\Lambda} \right) \, .
\end{equation}
Following the analysis in section (\ref{sec3}), the condition on the energy $\omega$ for having bound states is obtained simply making the replacement 
$G \rightarrow e^2/16 \pi \omega^2$ in  (\ref{w-grav-nonlocal}), namely:
\be
\om  > \frac{1}{2 S_\text{max} \, G \mu } 
%\quad \rightarrow \quad \om  > \frac{\omega}{2 S_\text{max} \, G \, \omega \, \mu  } 
\quad  \Longrightarrow \quad \boxed{\omega < \frac{\mu e^2 S_{\rm max}}{8 \pi} \approx \frac{e^2 M_{\rm P}}{8 \pi} }\, .
\ee
With again a similar analysis as the one in section (\ref{sec3}) and the replacement $\mu=1/\ell_\Lambda$, we get the following necessary and sufficient conditions for the existence of bound states, %($S_\text{max}/(8\pi)=0.53/(8\pi)\approx 0.021$),
\beq
\begin{split}
&
\boxed{\om <  0.02\mu e^2,
\qquad
U_\text{max} > \frac12,
\qquad
r_{b_1} < r_0 < r_{b_2}} \, ,
\\
&
\boxed{
0 < r_{b_1} <r_\text{min} < r_{b_2}  < r_\text{max},
\qquad
r_\text{min} < \frac{1.51}{\mu} < r_\text{max}<\frac{e^2}{8\pi\omega}}
\, . 
\label{constraint gs2}
\end{split}
\eeq

{\em We conclude that contrary to the previous examples, in nonlocal massless scalar electrodynamics the necessary condition for having bound states of pions is $\omega < e^2 \, M_{\rm P}$, and they have an extension $r_{b_2}$ that increases with increasing of the energy for fixed $r_0$ and $b$} . Moreover, the initial conditions must satisfy $b \ll a$ and $r_{b_1} < r_0 < r_{b_2}$ consistently with $t \ll s$.
{\em Since the existence of these bound states is based on a scalar field coupled to electromagnetism, we name these bound states: electroballs.}

%%%%%%%%%%%%%%%%%%%%%%%%%%%%%%%%%%%%%%%%%%%%%%%%%%%%%%%%%%%%%%%%%%%
%%%%%%%%%%%%%%%%%%%%%%%%%%%%%%%%%%%%%%%%%%%%%%%%%%%%%%%%%%%%%%%%%%%

\section{Local and nonlocal $\phi^3$ scalar field theories: Scalarballs}
\label{sec-scallar}
In this section, we study the issue of bound states in a class of local and nonlocal scalar field theories that have vanishing tree-level scattering amplitudes in the ultraviolet regime. In particular, we are here interested in a theory with positive dimension coupling $g$, contrary to the previous cases of negative dimension coupling or dimensionless couplings.

For a massless scalar with a cubic interaction, the general action reads:
\be
\hspace{-0.3cm}
S = 
  \int  d^D x \left[ \frac{1}{2} \,  \phi   f(\Box) \Box \phi - \frac{g}{3 !}  \phi^3  \right] ,
\label{scalarOnly}
\ee
where $f(\Box)$ can be a local or a nonlocal form factor. We here assume to recover the local two-derivative theory at large distances, namely $f(z) \rightarrow  1$ for $z \rightarrow 0$. 

It is straightforward to derive the tree-level scattering amplitude for $2 \rightarrow 2$ particles. The outcome in the $s$, $t$, and $u$ channels reads:
\be
 A_s = - \frac{g^2}{s f(s)}   \, , \qquad 
A_t = - \frac{ g^2 }{t  f(t)} \, , \qquad %\nonumber 
 A_u= - \frac{g^2 }{u f(u) }  \, . 
 \label{AmpliScalarL}
\ee
We can now compute the potential~\eq{red_pot} in $D=4$ taking the limit $t\ll s$, namely
\be
V(r) =- \frac{g^2}{16 \pi^2  \omega^2 r} \int_{0}^{\infty} \frac{d q}{q} \, \frac{\sin{(2 q r )}}{f(-q^2)}
\, .
\ee

Let us start with the two derivatives theory $f( - q^2) = 1$. In this case, we consistently obtain a Coulomb like potential, namely 
\be
V(r) =  -\frac{g^2}{32 \pi \om^2 r} \, .
\label{Vr}
\ee
Therefore, in the same fashion as in Einstein's gravity, bound states are not allowed for \eq{Vr} (see discussion at the end of Sec. \ref{sec2}). 
Indeed, the function~\eq{Q} for the potential ~\eq{Vr} is given by
\be
Q(r) = -1 + \frac{g^2}{64 \pi   \omega ^3 r}
,
\ee 
and the equation $Q(r) = 0$ has a single root at:
\be
\label{r_two}
r_\text{max}^\text{two-der} = \frac{g^2}{64 \pi  \omega ^3} \, ,
\ee
where the effective potential assumes the following maximum value, 
\be
U_\text{max} = \frac{1}{2}\left( \frac{32 \pi b \omega^3 }{e g^2} \right)^2 \, e^{\frac{ g^2}{32 \pi \omega^3 r_0} } \, 
.
\ee

%%%%%%%%%%%%%%%%%%%%%%%%%%%%%%%%%%%%%%%%%%%%%%%%%%%%%%%%%%%%%%%%%%%
%%%%%%%%%%%%%%%%%%%%%%%%%%%%%%%%%%%%%%%%%%%%%%%%%%%%%%%%%%%%%%%%%%% 

\subsection{Fourth derivative scalar theory} 
For the case of a local scalar higher derivative theory consistent with a soft amplitude in the ultraviolet regime (or softer than in the two-derivative theory),
we can take the form factor $f(\Box)$ to be polynomial. Thus, the simplest possible choice is:
\be
\hspace{-0.3cm}
f(\Box) = 1 - c_1 \Box %+ c_2 ( - \Box)^2.
\quad \rightarrow \quad f(-k^2) = 1 + c_1 k^2 \quad \rightarrow \quad f(t) = 1 - c_1 t 
\, ,
\label{Poly}
\ee
where $c_1$ is a parameter of inverse mass square dimension and is positive in order to avoid tachyons. 
Hence, for the above case the amplitude $A_t$ is not soft but actually divergent for $t = 1/c_1$. Moreover, the pole in $k^2 = -1/c_{1}$ corresponds to a ghost-like instability. 

If we do not care about such issues, the potential for the minimal local higher derivative theory (\ref{Poly}) is:
\be
V(r) =- \frac{g^2}{16 \pi^2  \omega^2 r} \int_{0}^{\infty} \frac{d q}{q} \, \frac{\sin{(2 q r )}}{(1 + c_1 q^2)}
= - \frac{g^2}{32 \pi  \omega^2} \frac{1 - e^{-2 \mu r} }{r},
\quad \text{where} \quad
\mu = \frac{1}{\sqrt{c_1}}\, . 
\label{4th_order_scalarpot}
\ee
The force is given by
\be
F(r) = -{ \frac{g^2 }{64 \pi  \omega ^2 r^2}}
\left(1 - e^{-2 \mu  r} - 2 \mu r e^{-2 \mu  r}  \right)
\, .
\label{4th_order_force}
\ee
Thus, we can prove that:
\beq
\begin{split}
&\lim_{r \to 0} V (r)
=-{\frac{g^2 \mu}{16\pi\omega^2} } \, ,
\qquad 
V (r)  \underset{r \to \infty}{\sim}  -\frac{g^2}{32 \pi \om^2 r}
,
\\
& \lim_{r \to 0} F (r)
= - {\frac{g^2 \mu^2}{32\pi\omega^2}} \, ,
\qquad 
F (r)   \underset{r \to \infty}{\sim} -{ \frac{g^2}{64 \pi \om^2 r^2}},
\end{split}
\label{4th force 0}
\eeq
and the conditions (i)-(iv) are satisfied.

Proceeding as in Sec.~\ref{sec3}, we look for the necessary and sufficient conditions for the existence of bound states. For the force~\eq{4th_order_force} the function~\eq{Q} has the following form, 
\be
Q(r) = -1 + \frac{g^2 \mu}{32\pi\omega^3} S( 2\mu r) \, ,
\label{K}
\ee
where the function $S(x)$ is now given by:
\be
S(x)=\frac{1}{x}[1-(1+x)e^{-x}].
\label{Definition A}
\ee
The function~\eq{K} has maximum 
\be
Q_\text{max} = - 1 + \frac{g^2 \mu}{32\pi\omega^3} S_\text{max} \, , 
\ee 
at the point $r = 1.79/2\mu$, where 
$
S_\text{max} = 0.30 
$ 
is the maximum value of~\eq{Definition A}. Therefore, $Q_\text{max}>0$ implies the following necessary condition for having a bound state, 
\be   
\omega <\Bigl(\frac{g^2 \mu \, S_\text{max}}{32\pi} \Bigr)^{1/3} \approx 0.14 \, (g^2 \mu )^{1/3} \, .
\label{condition omega s}
\ee
\begin{figure}[t]
\includegraphics[scale = 0.6]{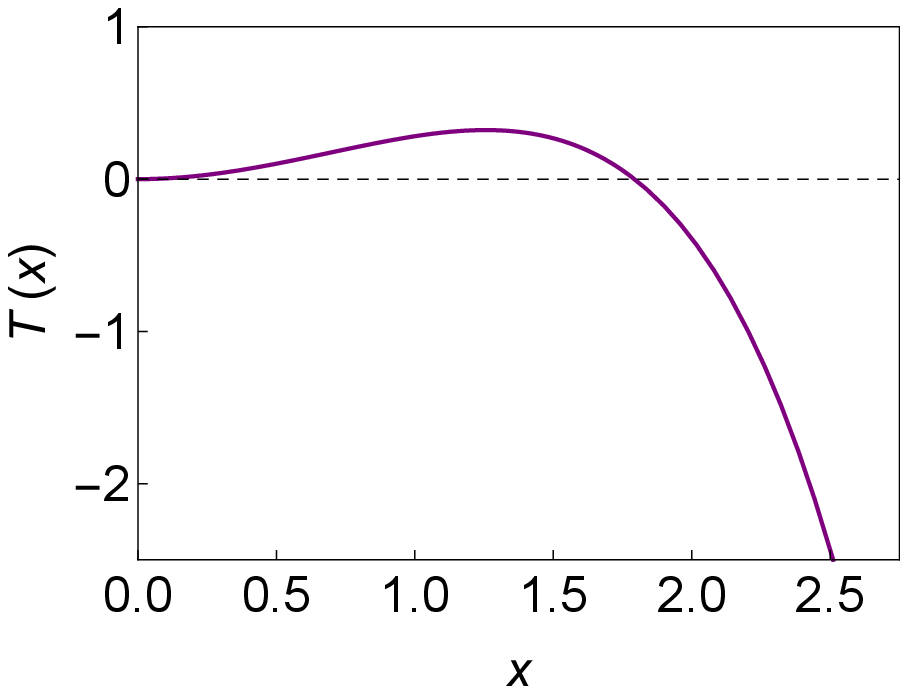}
\caption{\small Plot of the function~\eq{T4th} .
\label{T4thDerlabel}}
\end{figure}
The derivative of $Q$ at the equilibrium point $Q(r_e) = 0$ reads:
\be
\frac{dQ}{dr} \Big|_{r = r_e} = \frac{g^2 e^{-2\mu r_e}}{64\pi\omega^3 r_e^2} \, T(2\mu r_e) \, ,
\label{dQ4th}
\ee
where the function $T(x)$ is defined by
\be
T(x) = 1 + x + x^2 - e^x.
\label{T4th}
\ee
The feature of $T(x)$ is shown in Fig.~\ref{T4thDerlabel}. The equation $T(2\mu r_e)=0$ has a positive root at $r_e =1.79/2\mu$. For $0 <r_e <0.89/\mu$ we have $T( 2\mu r_e)>0$, while $T( 2\mu r_e)< 0$ if $r_e >0.89/\mu$. 
According to~\eq{d2W}, \eq{d2W2}, and~\eq{dQ4th} we find:
\be
0<r_\text{min}<\frac{0.89}{\mu}<r_\text{max} \, .
\label{C1}
\eea
The upper bound for $r_\text{max}$ is given by the value $r_\text{max}^\text{two-der}$ that we found in (\ref{r_two}) for the two-derivative theory. Indeed,
\be
\frac{\pa Q}{\pa \mu} = \frac{g^2 \mu r}{16\pi\omega^2} e^{-2\mu r} >0 \, ,
\ee
and the maximum value for $Q$ is obtained when $\mu \to \infty$. As a result, the necessary and sufficient conditions for the existence of scalarballs solutions in the fourth-derivative theory are:
\beq
\begin{split}
&
\boxed{\om < 0.14 \, (g^2 \mu )^{1/3} \, ,
\qquad
U_\text{max} > \frac12 \, ,
\qquad
r_{b_1} < r_0 < r_{b_2}} \, ,
\\
&
\boxed{0 < r_{b_1} <r_\text{min}  < r_{b_2} < r_\text{max} \, ,
\qquad
r_\text{min} < \frac{0.89}{\mu} < r_\text{max}  <  \frac{g^2}{64 \pi  \omega ^3}
\label{constraint g-4th}
}
\, .
\end{split}
\eeq
In Fig.~\ref{scalar_4} we show an example of bound states: in Fig.~\ref{scalar_4}~(a) we plot the effective potential~\eq{effective potential energy} and in Fig.~\ref{scalar_4}~(b) we have the corresponding numerical solution for the trajectory. 
\begin{figure}[t]
(a)\hspace{4cm} (b)\\
\includegraphics[width=4cm]{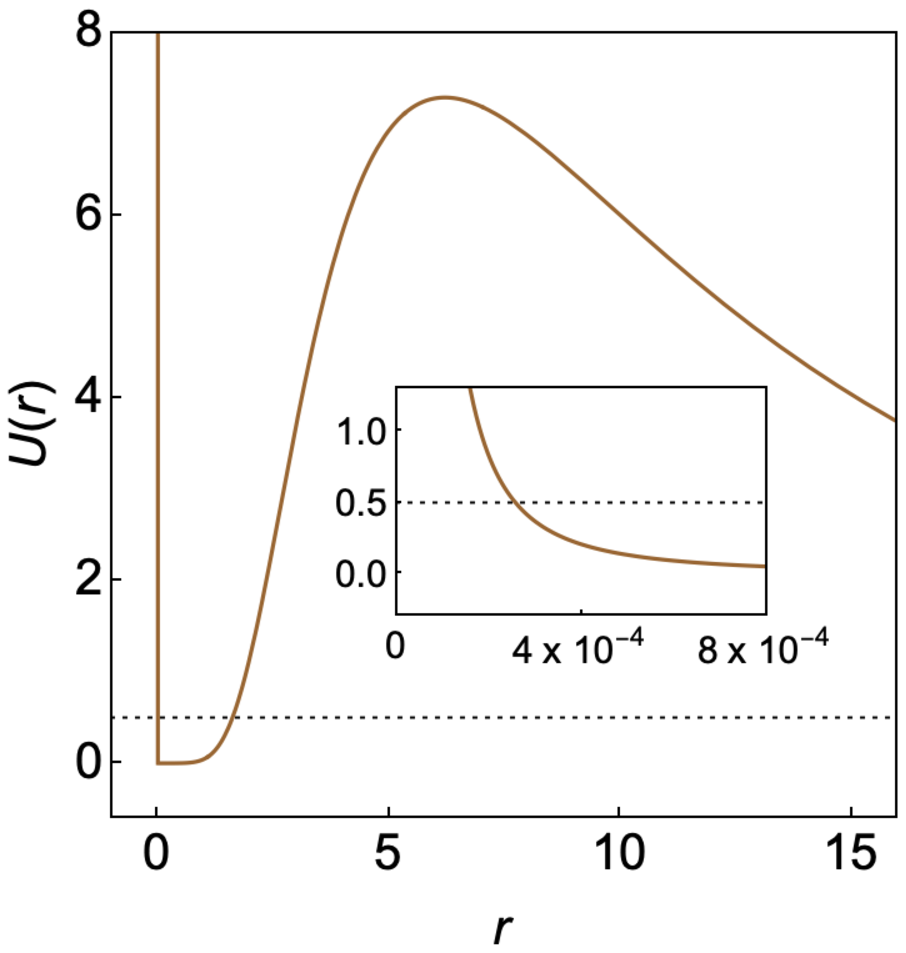}
\hspace{1cm}
\includegraphics[width=4.2cm]{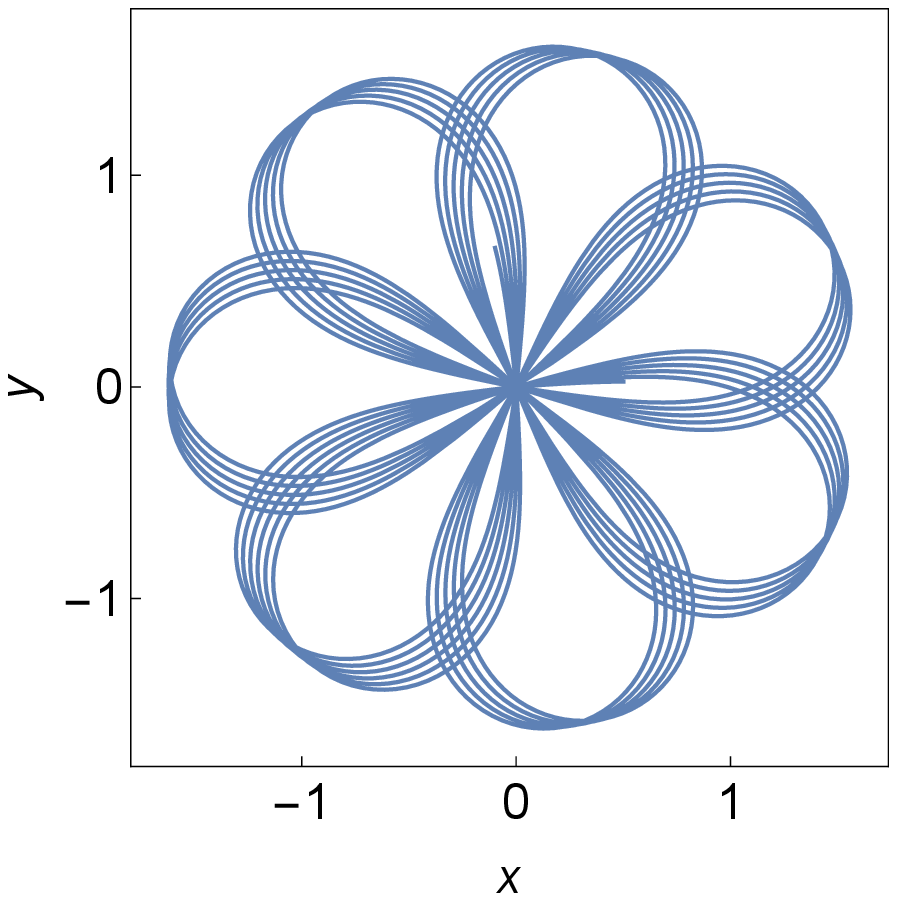}
\caption{\small Example of bound state for the fourth derivative scalar theory. Here we set $g=100$, $\mu=1$, $\omega=2$, $a=1$, $b=0.05$ and evolved the system till the final time $t_\text{f}=150$. (a) Effective potential. (b) The trajectory in the $xy$-plane.}
\label{scalar_4}
\end{figure}

%%%%%%%%%%%%%%%%%%%%%%%%%%%%%%%%%%%%%%%%%%%%%%%%%%%%%%%%%%%%%%%%%%%
%%%%%%%%%%%%%%%%%%%%%%%%%%%%%%%%%%%%%%%%%%%%%%%%%%%%%%%%%%%%%%%%%%% 

\subsection{Sixth derivative scalar theory}
In order to have an amplitude well defined for any real value of $s$, $t$, and $u$, 
we can choose the following quadratic polynomial:
\be
\hspace{-0.3cm}
f(\Box) = 1 + c_2 ( - \Box)^2 %+ c_2 ( - \Box)^2.
\quad \rightarrow \quad f(-k^2) = 1 + c_2 k^4 \quad \rightarrow \quad f(t) = 1 + c_2 t^2 
\, ,
\label{Poly2}
\ee
which corresponds to the minimal sixth-derivative theory with complex conjugate ghosts poles \cite{ShapiroModestoLW,ModestoLW}. 

For the form factor (\ref{Poly2}), the potential and the force are
\be
&& V(r) =  - \frac{g^2}{16 \pi^2  \omega^2 r}  \int \frac{dq}{q} \, \frac{\sin (2 q r)}{(1 + c_2 q^4)} 
= - \frac{g^2}{32 \pi  \omega^2}  \, \frac{1 - e^{- 2\mu r}  \cos \left( 2\mu r \right)}{ r}  
\,, \quad \text{where} \quad
\mu = \frac{1}{\sqrt{2} c_2^{1/4}} 
\, ,
\label{6th_order_scalarpot} \\
&&
F(r) =
-{\frac{g^2 }{64 \pi   \omega ^2 r^2}}
\left\{1 - [2 \mu  r \sin (2 \mu  r) + (1+ 2 \mu  r) \cos (2 \mu  r)] \, e^{-2 \mu  r} \right\} \, , 
\ee
and the conditions (i)-(iv) of Section \ref{sec2} read:
\beq
\begin{split}
&\lim_{r \to 0} V (r)
=-\frac{g^2 \mu}{16\pi\omega^2},
\qquad \qquad
V (r)  \underset{r \to \infty}{\sim}  - \frac{g^2}{32 \pi \om^2 r}
,
\\
& \lim_{r \to 0} F (r)
= 0,
\qquad \qquad \qquad \quad \,
F (r)   \underset{r \to \infty}{\sim} -{\frac{g^2}{64 \pi \om^2 r^2}}.
\end{split}
\eeq
The function $Q(r)$ is:
\bea
&& Q(r) =
-1 + \frac{g^2 \mu}{32\pi\omega^3} S(2 \mu r) \, ,
\label{Q-6th} \\
%\eea
&&
S(x)=\frac{1}{x}\bigl\{1-\bigl[\cos x+x(\cos x+\sin x)\bigr]e^{-x}\bigr\} \, .
\label{Defination H}
\ee
The above latter function has a maximum value $S_\text{max} = 0.46$ at $x = 2.06$, where~\eq{Q-6th} has also the maximum:
\be
Q_\text{max} = - 1+ \frac{g^2 \mu}{32\pi\omega^3} S_\text{max}
\, .
\ee
Therefore, the necessary condition for the existence of bound states, i.e. $Q_\text{max} >0$, implies the following inequality, 
\be
\omega<\Bigl(\frac{g^2 \mu \, S_\text{max}}{32\pi} \Bigr)^{1/3}
\approx 0.17 \, (g^2 \mu )^{1/3} \, .
\label{NS omega}
\ee
\begin{figure}[t]
\includegraphics[width=6cm]{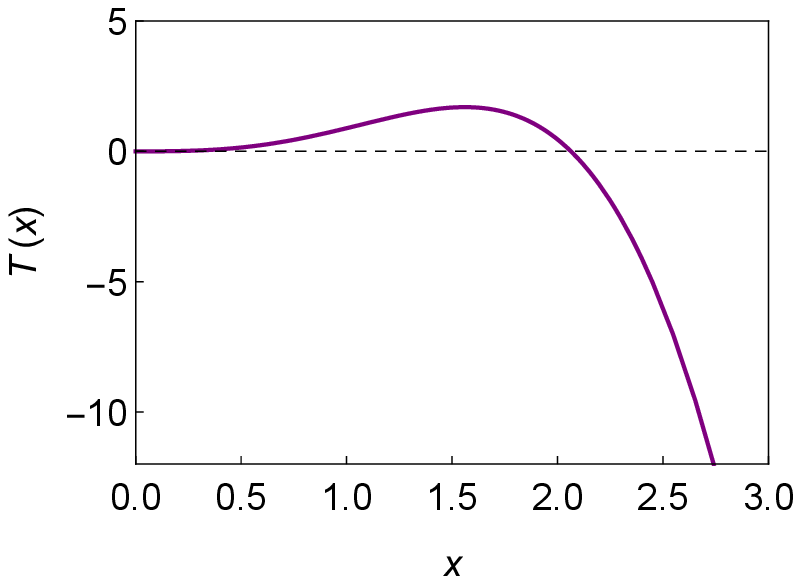}
\hspace{1cm} 
\includegraphics[scale=0.65]{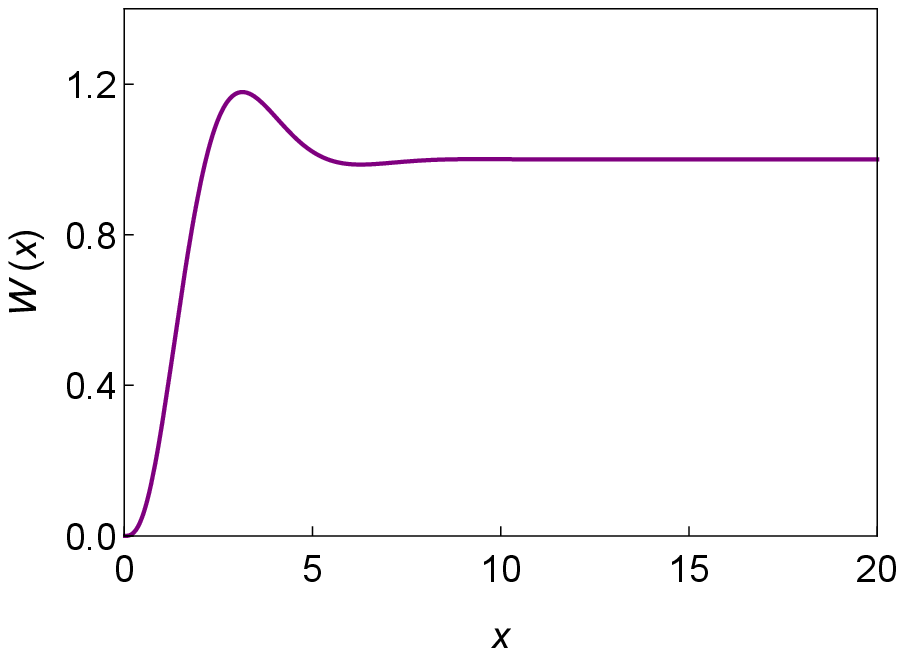}
\caption{\small Left panel: plot of the function~\eq{T6h}. Right panel: plot of~\eq{Defination H2}.\label{Alpha H}}
\end{figure}
Once again, the bounds on $r_\text{min}$ and $r_\text{max}$ can be derived computing the derivative of $Q(r)$, i.e. 
\be
\frac{d Q}{dr} \Big|_{r = r_e} = \frac{g^2 e^{-2\mu r_e}}{64\pi\omega^3 r_e^2} \, T(2\mu r_e) \, ,
\ee
\be
T(x) =(1+x)\cos x+x(1+2x)\sin x-e^x.
\label{T6h}
\ee
The plot of the function~\eq{T6h} is shown in Fig.~\ref{Alpha H}. Numerically we find that $T(2\mu r_e) = 0$ has a root at point~$r_{e} = 1.03/\mu$, so that $T(2 \mu r_e)$ is positive in the region $0 <r_e <1.03/\mu$, and it is negative for $r_e > 1.03/\mu$. Thus, the equilibrium points satisfy: 
\be
0<r_\text{min}<\frac{1.03}{\mu}
<r_\text{max}.
\label{rmin rmax}
\ee
Due to the oscillations of the potential \eq{6th_order_scalarpot}, $Q(r)$ {\it is not} monotonic with respect to $\mu$ for fixed $r$. Hence, the upper bound on $r_{\max}$ will be slightly bigger than for the two-derivatives~\eq{r_two}. In order to find an upper bound on $r_\text{max}$ we rewrite the function~\eq{Q-6th} as follows, 
\be
Q (r)
= - 1+\frac{g^2}{64\pi\omega^3 r} \, W(2 \mu r) \, ,
\label{S2}
\ee
where $W(x)$ is defined as:
\be
W(x)=1-\bigl[\cos x+x(\cos x+\sin x)\bigr]e^{-x} \, . 
\label{Defination H2}
\ee 
Since $W(x) \leqslant W_\text{max} = 1.18$ (see Fig. \ref{Alpha H}), $r_\text{max}$ is bounded by
\beq
- 1+\frac{g^2}{64\pi\omega^3 r_\text{max}} \, W_\text{max} > 0
\qquad \Longrightarrow \qquad
r_\text{max} < 1.18 \, \frac{g^2}{64\pi\omega^3 } 
.
\eeq
\begin{figure}[t]
(a)\hspace{4cm} (b)\\
\includegraphics[width=4.2cm]{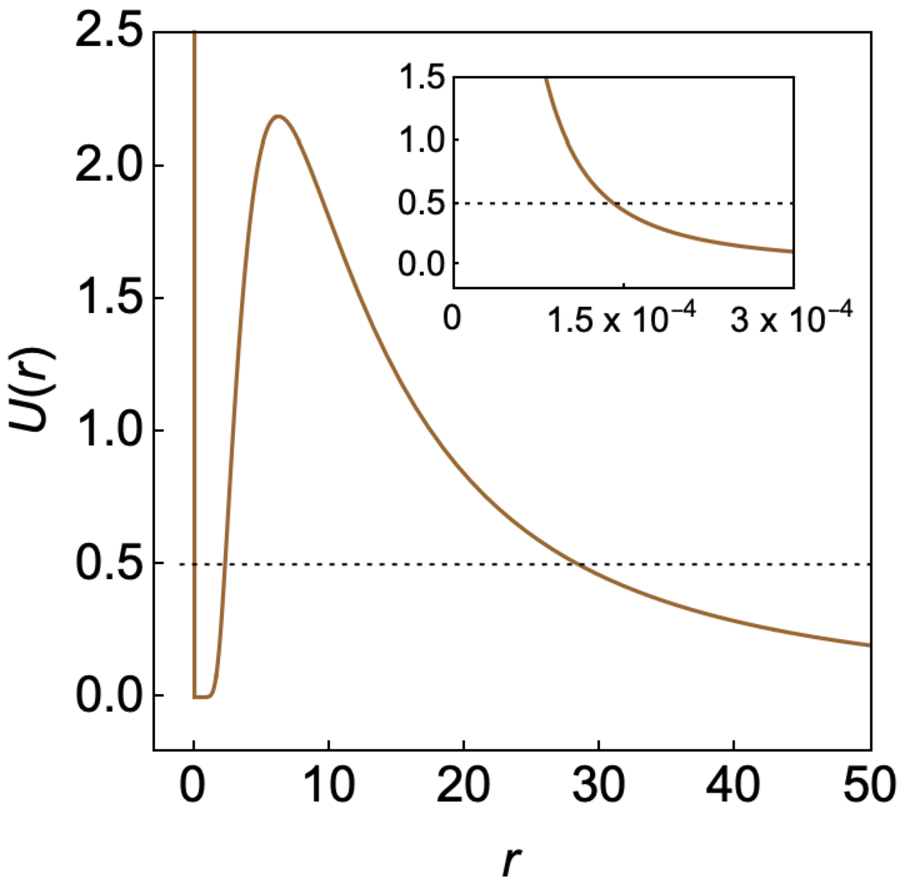}
\includegraphics[width=4cm]{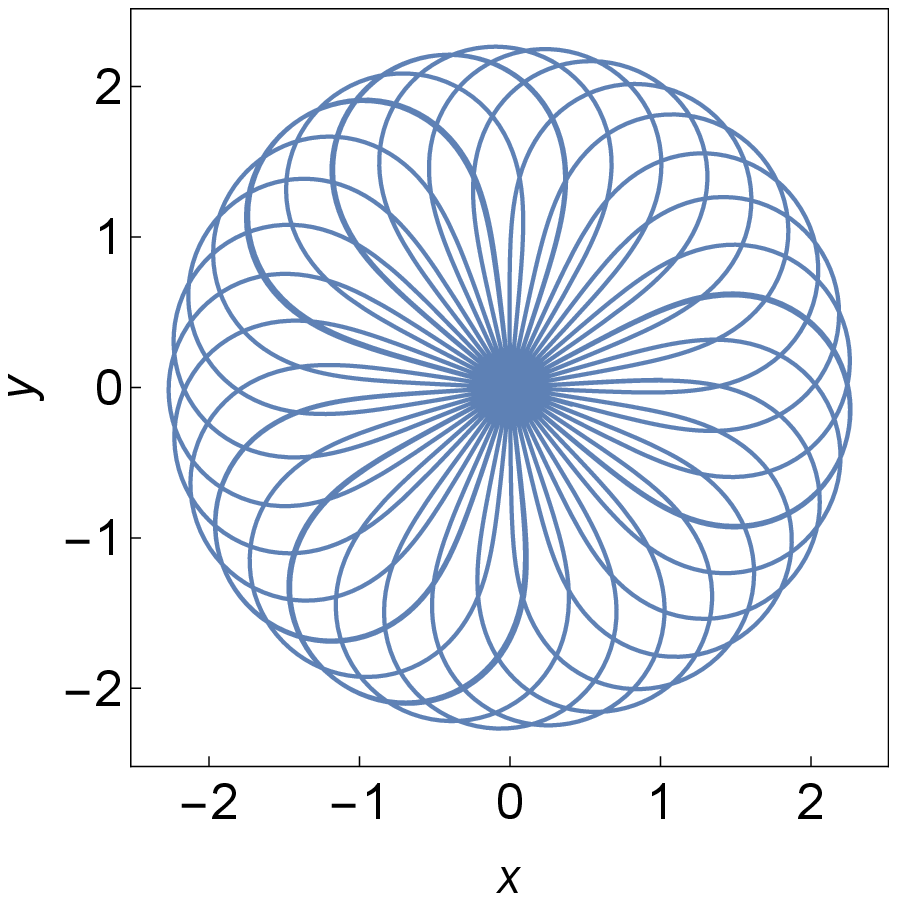}
\caption{\small  An example of bound state for the sixth derivative scalar theory. Here we set $g=100$, $\mu=1$, $\omega=2$, $a=2$, $b=0.1$ and evolved the system until a final time $t_\text{f}=200$: (a) Effective potential; (b) the trajectory of the system in the $xy$-plane.\label{S Trajectory 3}}
\end{figure}
Therefore, for the sixth-derivative theory the necessary and sufficient conditions for bound states are: 
\beq
\begin{split}
&
\boxed{\om < 0.17 \, (g^2 \mu )^{1/3},
\qquad
U_\text{max} > \frac12,
\qquad
r_{b_1} < r_0 < r_{b_2} } \, ,
\\
&
\boxed{ 0 < r_{b_1} <r_\text{min}  < r_{b_2} < r_\text{max},
\qquad
 r_\text{min} < \frac{1.03}{\mu} < r_\text{max}< 1.18 \, \frac{g^2}{64 \pi  \omega ^3}
 } 
 \, .
\label{constraint g-6th}
\end{split}
\eeq
Finally, in Fig.~\ref{S Trajectory 3} we show an example of scalarball for a set of parameters that satisfy~\eq{constraint g-6th}.

%%%%%%%%%%%%%%%%%%%%%%%%%%%%%%%%%%%%%%%%%%%%%%%%%%%%%%%%%%%%%%%%%%%
%%%%%%%%%%%%%%%%%%%%%%%%%%%%%%%%%%%%%%%%%%%%%%%%%%%%%%%%%%%%%%%%%%% 

\subsection{Nonlocal scalar field theory}

We hereby extend the result of the previous sections to a nonlocal scalar theory. For this purpose, we simply have to replace $f(\Box)$ with an analytic function without zeros in the all complex plane at finite distance, namely 
\be 
f(\Box) = e^{H(\Box)} \, ,
\ee
and the amplitude (\ref{AmpliScalarL}) turns into:
\be
 A_s = - \frac{g^2}{s} e^{- H(s)} \, , \quad 
A_t = - \frac{ g^2 }{t} e^{- H(t)} \, , \quad 
 A_u= - \frac{g^2}{u} e^{- H(u)} \, . 
 \label{AmpliScalarNL}
\ee

\begin{figure}[t]
(a)\hspace{4cm} (b)\\
\includegraphics[width=4.2cm]{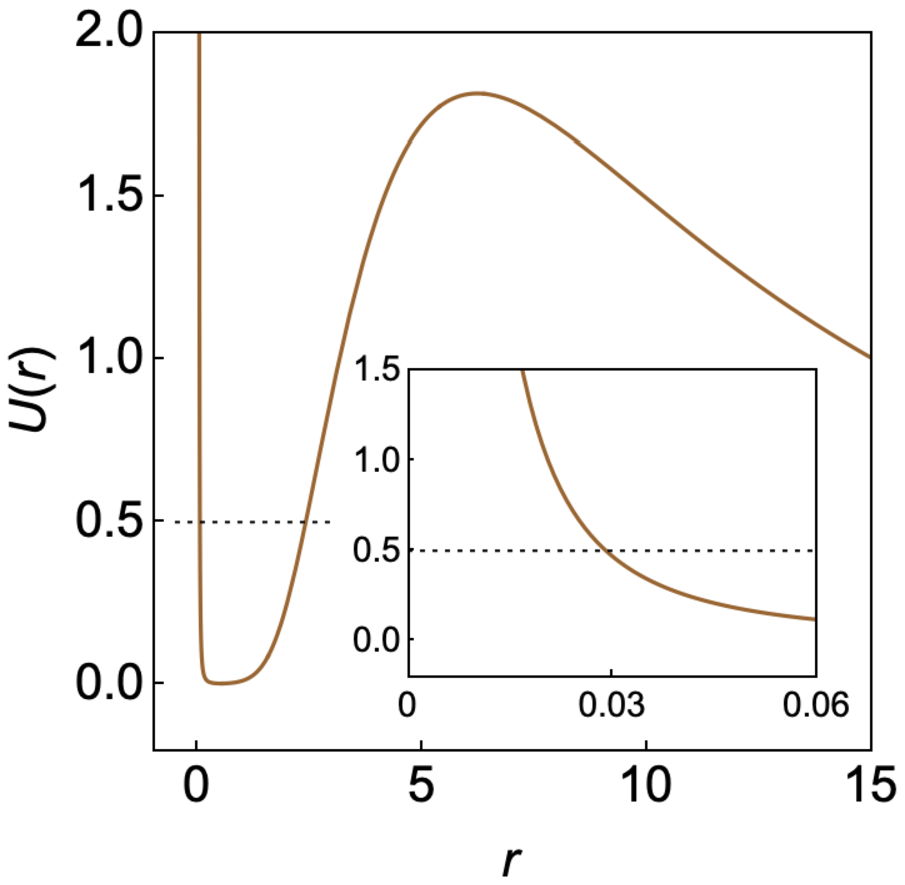}
\includegraphics[width=4cm]{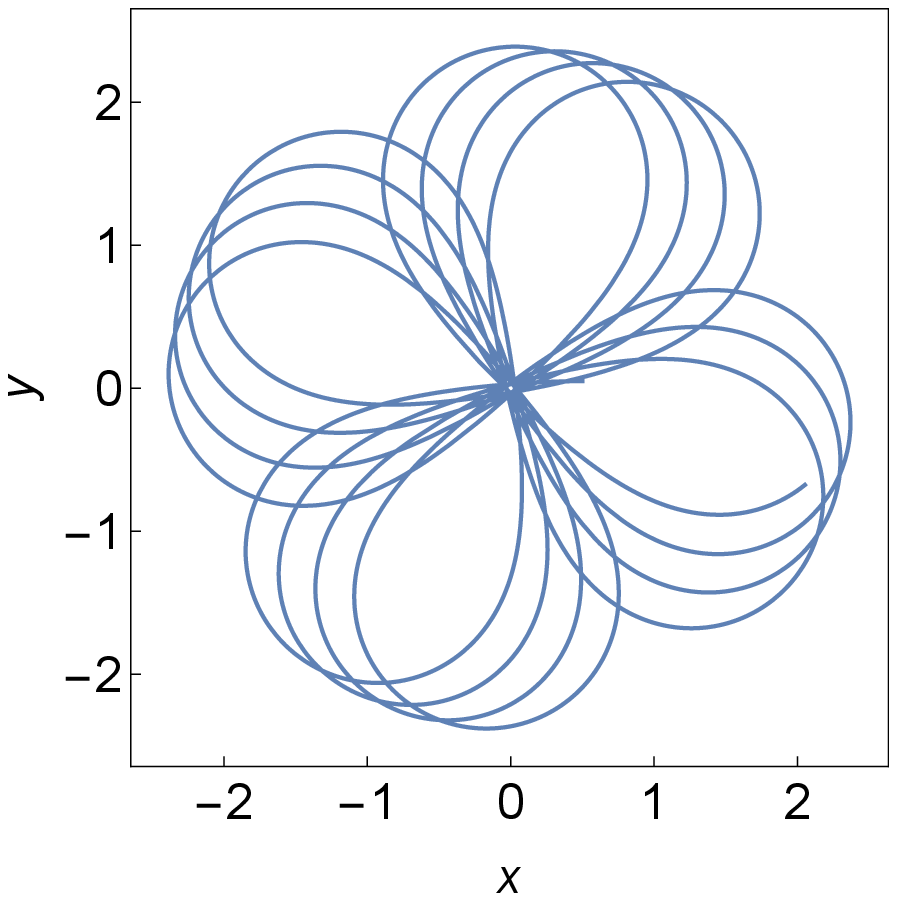}
\caption{\small An example of a bound state for the nonlocal scalar theory with $g=100$, $\mu=1$, $\omega=2$, $a=1$, $b=0.1$, and $t_\text{f} = 100$: (a) Effective potential; (b) the trajectory of the system in the $xy$-plane.\label{Bound state NS}}
\end{figure}

As example, we take the form factor $e^{-\si \Box}$. As discussed before, the amplitude, in this case, is not soft. However, if we again do not care about such problem, the potential \eq{red_pot} after integration reads:
\bea
V(r)
= -\frac{g^2}{32\pi \om^2 r} \, \text{erf} (\mu r ), 
\qquad \qquad \mu = \frac{1}{\sqrt{\si}}.
\label{Scalar Nonlocal 1}
\eea
Since the potential~\eq{Scalar Nonlocal 1} has the same functional form as the one considered in Sec.~\ref{sec3}, many of the intermediate steps are the same as Sec.~\ref{sec3}. However, the final conclusion for the bound on the energy $\om$ is in contrast with~\eq{w-grav-nonlocal-2} because of the different front coefficient in~\eq{Scalar Nonlocal 1}. So, in order to avoid repetitions, we refer to Sec.~\ref{sec3}, and we here just present the necessary and sufficient conditions for having bound states in the nonlocal scalar theory, namely 
\beq
\begin{split}
&
\boxed{\om < 0.14 \, (g^2 \mu )^{1/3},
\qquad
U_\text{max} > \frac12,
\qquad
r_{b_1} < r_0 < r_{b_2} } \, ,
\\
&
\boxed{
0 < r_{b_1} <r_\text{min}  < r_{b_2} < r_\text{max},
\qquad
r_\text{min} < \frac{1.51}{\mu} < r_\text{max} <  \frac{g^2}{64 \pi  \omega ^3}
}
\, .
\label{gmass}
\end{split}
\eeq
In Fig.~\ref{Bound state NS} we show an example of scalarballs for the nonlocal scalar field theory. 

{\em We conclude that in the $\phi^3$ higher derivative or nonlocal theory,  a necessary condition for bound states of scalars is $\omega < M_{\rm P}$,\footnote{For the sake of simplicity we here assume $g = \mu = M_{\rm P}$.} }
{\em and extension $r_{b_2}$ increases for increasing energy for fixed $r_0$ and $b$}. Moreover, the initial conditions must satisfy $b \ll a$.

%%%%%%%%%%%%%%%%%%%%%%%%%%%%%%%%%%%%%%%%%%%%%%%%%%%%%%%%%%%%%%%%%%%
%%%%%%%%%%%%%%%%%%%%%%%%%%%%%%%%%%%%%%%%%%%%%%%%%%%%%%%%%%%%%%%%%%%

\section{Asymptotically free or finite theories: Planckballs} \label{Planckballs}

In force of the results in the previous section, we can make a proposal for scattering amplitudes consistent with perturbative bound states at short distances and asymptotic freedom.  
 As a particular example, we can consider the following proposal for the scattering amplitude of two-gravitons into two-gravitons including the loop-corrections, namely 
  \be
% A_s =  8 \pi G \frac{t^3}{s u F(s)}  \, , \quad 
A (+,+; +,+)=  8 \pi G \frac{s^3 }{t u} \, \frac{1}{f(t)}  \, ,  
% A_u= - 8 \pi G \frac{u^3}{s u F(u)}  \, . 
 \label{Amplitudes2}
\ee
where the $+$ stays for the helicity of the graviton and $f(t)$ now can only come from the loop corrections. Indeed, in well defined nonlocal theories~\cite{Modesto:2021okr, Modesto:2021ief} 
the tree-level amplitudes are the same as the local standard model coupled to gravity \cite{MeSpacetime-Matter} so that the classical limit is recovered for $f(t)=1$. 
Let us shortly remind the theory in~\cite{Modesto:2021okr, Modesto:2021ief,MeSpacetime-Matter}, 
\be
&& S[\Phi_i] = \int {\rm d}^D x \sqrt{-g} \left( {\mathcal L}_\ell + E_i \, F(\Delta)_{ij} \, E_j \right) \, , 
\label{action} \\
&& S_{\ell} = \int {\rm d}^D x \sqrt{-g} \, {\mathcal L}_\ell \, ,\quad 
 \mathcal{L}_\ell = \frac{2}{\kappa^2} R + \mathcal{L}_m \, , 
  \label{localT}\\
&& E_i(x) =  \frac{\delta S_\ell }{\delta \Phi_i(x)} \, , 
\label{EH} \\
&& 2 \Delta_{i k} F(\Delta)_{k j} \equiv \left(e^{H( \Delta_\Lambda)} - 1 \right)_{i j},
\label{FF}
\ee
where $\Phi_i$ is a set of fields including the spacetime metric, $F(\Delta)_{ij}(x,y)$ is a symmetric tensorial entire function whose argument is the Hessian operator $\Delta$, and $H(\Delta)$ is an entire analytic function. $\Lambda$ is the non-locality scale. 

Therefore, in the nonlocal unified theory of gravity and matter, the bound states can only form because of the quantum corrections that make the scattering amplitude smooth in the ultraviolet regime.
We are at the moment investigating such amplitudes, but we already have indications about the scaling of the quantum form factor at high energy. Indeed, the quantum corrections after resummation of the one-loop one-particle irreducible contributions lead to a form factor with the following ultraviolet polynomial scaling,
\be
f(\Box) \sim \Box^{n} \, ,
\ee
where $n>3$ for $D=4$. 
However, in general, it is a hard task to determine the quantum form factor $f(\Box)$; hence, we limit here to comment on what should be the outcome of the loop-computations and resummation in order to have bound states at high energy. 
The limit $s \gg t$ for the amplitude (\ref{Amplitudes2}) reads:
 \be
% A_s =  8 \pi G \frac{t^3}{s u F(s)}  \, , \quad 
A (+,+; +,+) \approx  - 8 \pi G \frac{s^2 }{t} \, \frac{1}{f(t)}  \, . \quad %\nonumber 
% A_u= - 8 \pi G \frac{u^3}{s u F(u)}  \, . 
 \label{sAmplitudes2}
\ee
Therefore, the interaction potential~\eq{red_pot} is:
\beq
V(r) =  - \frac{8  G \omega^2}{ \pi \, r} \int_{0}^{\infty} d q \, \frac{\sin{(2  r q )}}{q f(-q^2/\Lambda^2)}
\, .
\label{pot_general}
\eeq
As discussed in Sec.~\ref{sec2}, the potential should satisfy the conditions (i)-(ii) for the existence of bound states. The convergence properties of the integral~\eq{pot_general} with general form factor $f(-q^2/\La^2)$ have been studied exhaustively in~\cite{Giacchini:2018wlf,Burzilla:2020utr}. The main outcome is that $V(r)$ is finite at $r=0$ if the form factor grows as $f(t) \sim t$ or faster for sufficiently large values of the argument $t$. On the other hand, $V(r) \sim 1/r$ for large $r$ consistently with $f(t) \to 1$ when $t\to 0$. 

Therefore, {\em the minimal outcome for the quantum form factor $f(\Box)$ in order to have a bound state is: $f(\Box) \sim \Box$ for large $\Box$}. Given the preliminary results mentioned above, bound states seem very likely to form at high energy.

Since the result is based on a nonlocal ultraviolet completion of all fundamental interactions, let us call these bound states generically: {\em Planckballs.} In the latter class of bound states, the suitable name for the states resulting from the amplitude (\ref{sAmplitudes2}) is {\em quantum graviballs}. 
The properties of the Planckballs are the same as the gravi-scalarballs, namely they form for $\omega \gtrsim M_{\rm P}$ and they have an extension that decreases with the increase of the energy\footnote{We here assume the non-locality scale $\Lambda$ to be the Planck mass.} when the other parameters $r_0$ and $b$ are fixed. Moreover, the initial conditions must satisfy $b \ll a$ consistently with $t \ll s$.

If we consider a scattering process mediated by an interaction governed by a dimensionless coupling constant $g$, then the outcome is the same of section (\ref{Electroballs}), namely bound states form if $\omega \lesssim g^2 M_{\rm P}$. 
Let us expand on the latter statement considering the simple example of the nonlocal scalar electrodynamics not defined like in section (\ref{Electroballs}) but using the recipe (\ref{FF}). The theory is explicitly obtained replacing the following local Lagrangian in (\ref{FF}), 
\be
\mathcal{L}_{\ell} =  - \frac{1}{4 e^2} F_{\mu\nu} F^{\mu\nu} 
- (D_\mu \Phi)^{\dagger} (D^\mu \Phi)\, , 
%\quad 
%\sum_{a}^{N_f}\bar \psi_a \, i \slashed{\cal D}  e^{H(-\slashed{\cal D}^{2}_{\Lambda})}\, \psi_a
\label{LED}
\ee
Therefore, the tree-level amplitude is the same of the local theory, namely 
\be
%\mathcal{M}^{\text{tree}}=
A_{t}^{\rm tree} \approx - 2 i e^2 \frac{s}{t} \, .
\ee
However, after resummation, according to the discussion above in this section, we expect:
\be
A_t^{\rm 1PI} \approx  - 2 i e^2 \frac{s }{t} \, \frac{1}{f(t)}  \, . 
 \label{sAmplitudes3}
\ee
Thus, the results of section (\ref{Electroballs}) can be applied to the amplitude (\ref{sAmplitudes3}) for the class of quantum corrections discussed above.

%%%%%%%%%%%%%%%%%%%%%%%%%%%%%%%%%%%%%%%%%%%%%%%%%%%%%%%%%%%%%%%%%%%
%%%%%%%%%%%%%%%%%%%%%%%%%%%%%%%%%%%%%%%%%%%%%%%%%%%%%%%%%%%%%%%%%%%  

\section{Summary and Discussion}

We have shown that bound states of massless particles are allowed in: (I) a sterile scalar field theory coupled to local or nonlocal higher derivative gravity, (II) string theory, (III) nonlocal scalar electrodynamics, (IV) a $\phi^3$ higher derivative or nonlocal field theory, and (V) general asymptotically free or finite NLQG and HDQG. 
This is mainly due to the weakness of the fundamental interactions at short distances and, in particular, to the absence of singularity in the potential $V(r)$ at $r=0$. The latter one is a universal property common to all asymptotically free or finite theories. 

In order to prove the existence of bound states and provide explicit examples, we used the dynamical framework proposed  in~\cite{Guiot2020} that consists of a set of equations of motion for relativistic massless particles interacting through a potential defined as the Fourier transform of the scattering amplitude in the Regge's limit. 
From the technical point of view, using polar coordinates, we reduced the dynamical system to a one-dimensional problem of a particle moving in an ``effective potential'' $U(r)$.
By studying such effective potential, we proved that we could not have bound states in two derivatives' theories and, in particular, in Einstein's gravity. 
On the other hand, in nonlocal gravity and string theory, the correct bounds on the particles energy $\omega$ are given respectively by~\eq{w-grav-nonlocal} and~\eq{string-1}--\eq{string-2}. 
However, for the sake of simplicity, we can assume that the fundamental scales $\ell_\Lambda$ and $\sqrt{\alpha'}$ are both equal to the Planck length $\ell_{\rm P}$. Therefore, the only fundamental scale is the Planck mass $M_{\rm P}$. 
In order to study the feasibility of bound states, the complete set of initial parameters consists of:
\be
&&
 \omega : \quad \mbox{the energy of the massless fundamental particles} \, , \nonumber \\
&&
b  : \quad \mbox{impact parameter} \, , \nonumber \\
&& 
r_0 \equiv r(t_0) = \frac{\sqrt{a^2+b^2}}{2} : \quad \mbox{half of the initial distance between the two particles} \, ,
\ee
which must satisfy the following conditions:
\be
b \ll a \, ,  \quad r_{b_1} < r_0 < r_{b_2} \, , 
\label{InCon}
\ee
consistently with $t \ll s$. In (\ref{InCon}) $r_{b_1}$ and $r_{b_2}$ are respectively the two turning points of the orbital radius. 

Hence, the {\em necessary} condition for having {\em bound states of scalars or gravitons} respectively in the theories \underline{(I) and (II)} reads:
\be
 \omega > M_\text{P} \quad \Longrightarrow \quad \mbox{bound states}
 \, .
 \label{omegaMP}
\ee

$\bullet$ If we keep {\bf fixed $r_0$ and $b$ (and $a$ also), and we vary the energy $\omega$}, it turns out that the bound states extend to the maximum orbital radius $r_{b_2}$ that decreases with increasing the energy $\omega$.
Therefore, very massive bound states have a very small size despite having a very large mass.

$\bullet$ If we {\bf decrease $b$ or $r_0$ keeping fixed the energy $\omega$} (and $a$ also), then $r_{b_2}$ increases (see Appendix.~\ref{extension}).

For the theories with \underline{(III) and (IV)}, 
the conditions for having ``electroballs'' and ``scalarballs'' are opposite to those in nonlocal gravity and string theory. Since $e$ is dimensionless and $g$ has dimension of mass, the condition for creating a bound state is $\omega < M_\text{P}$. Moreover, if $\omega$ increases, both $r_{b_1}$ and $r_{b_2}$ increase. In these theories, very light bound states have a small extension. 
Indeed, 
$r_{b_2}$ decreases if $\omega$ decreases for fixed $a$ and $b$, although $r_\text{max}$ increases.

Summarizing even further: {\em the necessary condition for having bound states depends on the dimension of the coupling constant. If the coupling constant has mass dimension less than $-1/2$ the condition is $\omega > M_\text{P}$, if the coupling constant has dimension larger than $ - 1/2$ then the condition turns into  $\omega < M_\text{P}$}\footnote{Let us assume $[g]\sim M^n$ and write the potential as $V(r)\sim -g^2 \omega^{-2n} r^{-1} W(\mu r)$. Therefore, we have:
\be 
Q(r)=-1-r\omega^{-1} F(r) \sim -1+ \mu g^2 \omega^{-(2n+1)} S(\mu r) \, . 
\nonumber
\ee
Here both $W(\mu r)$ and $S(\mu r)$ are dimensionless functions. Since the necessary condition for having bound states is given by $Q_\text{max}>0$, we get:
\be \omega^{2n+1}<S_\text{max} g^2 \mu \, . 
\nonumber
\ee
Therefore, if $n > - 1/2$ (namely $2n+1>0$) we find $\omega < M_\text{P}$.}.

In the general unified theory (V), we explicitly considered the case of graviballs. However, we expect the same results listed above to apply according to the mass dimension of the coupling constant involved in the scattering process. 
The asymptotic freedom \cite{Briscese:2019twl, Rachwal:2021bgb} in super-renormalizable theories \cite{Krasnikov, Modesto, ModestoLeslawR} or the softness of the quantum amplitudes in finite theories \cite{Kuzmin, ModestoLeslawF, Modesto:2021okr, Modesto:2021ief} guarantee the stability and Universality of our result in a consistent theory of all fundamental interactions~\cite{Modesto:2021okr, Modesto:2021ief,MeSpacetime-Matter}.

From the geometric point of view, for scattering amplitudes mediated by the graviton field, we can infer about the spacetime metric.
For the potential $V(r)$ (\ref{deltaSFT}), we can here provide the explicit example of metric, which in isotropic coordinates is given by
\be
\label{metric1}
ds^2 \approx - \left( 1 + \frac{2 V(r)}{\omega} \right) dt^2 + \left( 1 - \frac{2 V(r)}{\omega} \right) ( dr^2 + r^2 d \Omega^{2} ) 
\, ,
\quad \quad 
\left| 2 V(r) / \omega \right| \ll 1.
\ee
By changing to the Schwarzschild coordinates ($t,R,\theta,\ph$), the metric \eq{metric1} boils down to
\be
\label{metric2}
ds^2 \approx - \left( 1 + \frac{2 V(R)}{\omega} \right) dt^2 + \left( 1 + \frac{2 R V'(R)}{\omega} \right)  dR^2 + R^2 d \Omega^{2}  
\,,
\ee
where here the ``prime'' denotes diferentiation with rescpect to $R$. The metric~\eq{metric2} is singularity free with a de~Sitter core in $R=0$. Given a metric in the form~\eq{metric2}, the event horizon is defined as the solution of the equation $(\na R)^2 = 0$~\cite{FNBook}. For the potential (\ref{deltaSFT}) we get
\be
(\na R)^2 = 0 \quad \Longrightarrow \quad  \frac{16 G\omega \mu}{\sqrt{\pi}} \, e^{-(\mu R)^2} \simeq - 1 
\, .
\ee
Since the above equation has no solutions, the metric \eq{metric2} does not have an event horizon. 
Similar horizonless exotic compact objects (named $2$-$2$-holes remnants 
\cite{Aydemir:2020xfd, Aydemir:2020pao, Holdom:2022zzo}) where discovered in Stelle's Gravity and proposed as candidate dark matter. Such $2$-$2$-holes may be viewed as non-perturbative bound states of gravitons and massive spin-$2$ ghost-like states in quadratic gravity \cite{Holdom:2015kbf, Holdom:2019ouz}. 

In the main text we proved that the maximum value for $r_{b_2}$ is $\approx 2 G \om$ for large $\omega$. However, even though the bound states for gravitons and particles interacting gravitationally are confined inside the Schwarzschild radius, the metric~\eq{metric2} does not have an event horizon because of the weakness of the interactions that survive at the nonlinear level as a feature of asymptotic freedom or finiteness. 
Other spacetimes with similar properties can be obtained for the potentials (\ref{n2pot}) and~(\ref{stringPote}).
The result in (\ref{metric1}) is in agreement with the spacetime found in \cite{Nicolini:2005vd,Modesto:2010uh}, where for a value of the mass close to the Planck mass, the solution does not show either the event horizon or the Cauchy' horizon. 
On the other hand, for large $\omega$, we can have:
\be
\left| 2 V(r) / \omega \right| =  \frac{4 G \om}{r} \, {\rm erf}\left( \frac{r}{\ell_\Lambda} \right) \gtrsim 1 \quad {\rm or} \quad \gg 1 \, .
\ee
because ${\rm erf}\left({r}/{\ell_\Lambda} \right)/r$ is a limited function. 
Hence, we enter in the nonlinear regime, and the metric at large distances must reproduce the Schwarzschild one according to Einstein's gravity, while at short distances, the asymptotic freedom of the theory will guarantee the correctness of (\ref{metric1}) also at the nonlinear level. The outcome is a black hole metric similar to the one in  \cite{Nicolini:2005vd,Modesto:2010uh, Bambi:2016uda, Zhang:2014bea, Bambi:2013gva} or \cite{Buoninfante:2019swn} depending on whether we impose or not the condition $g_{tt} = - g_{RR}^{-1}$.

On the other hand, for scattering amplitudes that do not involve gravity but Abelian gauge fields --- coupled to matter --- and non-Abelian gauge fields (coupled or not to matter), we can claim {\em perturbative confinement}. Notice that in higher derivative local or nonlocal non-Abelian gauge theories, we have two confining phases: a perturbative one in the ultraviolet regime and a non-perturbative one in the infrared regime. This is a feature of the higher derivative or nonlocal theories if the quantum amplitudes at high energy fall off faster than $1/(t \log t)$ as shown in section (\ref{Planckballs}).
It turns out that the condition for having bound states in gauge theory is opposite to the one involving gravity, namely $\omega < M_{\rm P}$. The latter statement has been derived for the toy models (III) and (IV), and it is mainly due to the zero or positive dimension of the coupling constants. 

Finally, in the unified standard model theory (V), both possibilities $\omega > M_{\rm P}$ or $\omega < M_{\rm P}$ may arise depending on the dimension of the coupling constant and according to the examples (I), (II), or (III).

We end up with a comment on the creation of bound states in the early Universe. 
If the energy density for matter is $\rho$, $[\rho] = \rm{M/L^3}$, and $\langle E \rangle$ is the average energy per particle, the number density is:
\be
\rho_{\rm N} = \frac{\rho}{\langle E \rangle} \, , \quad [ \rho_{\rm N}] = \frac{\mbox{Number of particles}}{L^3}\, .
\ee
We can also define the linear density, namely 
\be
\rho_{\rm N, \ell} = \left( \frac{\rho}{\langle E \rangle} \right)^{\frac{1}{3}} \, , \quad [ \rho_{\rm N, \ell}] = \frac{\mbox{Number of particles}}{L}\, .
\ee
Therefore, the average free path is: 
\be
\langle d \rangle = \frac{1}{\rho_{\rm N,\ell}} = \left( \frac{ \langle E \rangle }{ \rho }\right)^{\frac{1}{3}} \, .
\ee
Since $\langle d \rangle \sim r_0$ and $r_0 \lesssim r_{b_2}$, we get the following condition, 
\be
r_0 \lesssim r_{b_2} \quad \Longrightarrow \quad r_{b_2} \gtrsim \left( \frac{\langle E \rangle}{M_{\rm P} } \frac{M_{\rm P}}{\rho} \right)^{\frac{1}{3}} 
\quad \Longrightarrow \quad \frac{ \langle E \rangle}{M_{\rm P}} \lesssim \frac{\rho}{M_{\rm P} } r_{b_2}^3 \, .
\label{rb2c}
\ee
For $\omega \sim \langle E \rangle > M_{\rm P}$ (\ref{rb2c}) simplifies to:
\be
r_{b_2} \gtrsim \left( \frac{M_{\rm P}}{\rho} \right)^{\frac{1}{3}} \, .
\label{rb2cd}
\ee
Using the conserved quantity $\rho a^n = \rho_0 a_0^n$ ($n=4$ for radiation and $n=3$ for dust matter),  (\ref{rb2cd}) turns into:
\be
r_{b_2} \gtrsim \left[ \frac{M_{\rm P}}{\rho_0}  \left( \frac{a}{a_0} \right)^n \right]^{\frac{1}{3}} \, .
\label{rb2cde}
\ee
For $n=4$ (radiation), $a \sim t^{1/2}$, $t=t_{\rm P}$, $t_0 = 10^{17}$ sec, 
$\rho_0 = 1.8 \times 10^{-29} {\rm gr}/{\rm cm}^3 = 1.8 \times 10^{-24} M_{\rm P}/{\rm cm}^3$ (where $M_{\rm P} = 10^{-5} {\rm gr}$), 
\be
r_{b_2} \gtrsim \left[ \frac{M_{\rm P}}{\rho_0}  \left( \frac{t_{\rm P}}{t_0} \right)^2 \right]^{\frac{1}{3}} \approx \ell_{\rm P}
\, ,
\label{rb2cdef}
\ee
which is consistent with $r_{b_2} < r_{\max} < 2 G \omega = 2 G \langle E \rangle = 2 \ell_{\rm P}$ 
(for $\langle E \rangle = M_{\rm P}$) in (\ref{constraint gs}). 
If we go even further back in time, $a(t)$ decreases together with the lower bound on $r_{b_2}$. 
Hence, we can speculate that the bound states in the theories (I) and (II) are very likely formed in the early Universe and can be a proposal for a dark matter candidate, or simply, to be the seeds for the later structure formation. 
Notice that if the bound states are formed in the Planckian or Trans-Planckian regime, they will be diluted by inflation. 
Therefore, we need to replace inflation with another mechanism to solve the problems of classical cosmology and/or to fit the CMB spectrum for the scalar and tensor density perturbations.

For $\omega < M_{\rm P}$, for example $\omega = \langle E \rangle = 10^{-5} M_{\rm P}$, 
$n=4$ (radiation), $a \sim t^{1/2}$, $t=t_{\rm inflation} \sim 10^{-31}$ sec, $t_0 = 10^{17}$ sec, 
$\rho_0 = 1.8 \times 10^{-29} {\rm gr}/{\rm cm}^3 = 1.8 \times 10^{-24} M_{\rm P}/{\rm cm}^3$ ($M_{\rm P} = 10^{-5} {\rm gr}$), 
 the inequality (\ref{rb2c}) reads:
\be
r_{b_2} \gtrsim \left[ \frac{10^{-5} M_{\rm P}}{\rho_0}  \left( \frac{t_{\rm inflation}}{t_0} \right)^2 \right]^{\frac{1}{3}} \approx 8 \times 10^8 \, \ell_{\rm P}
\, ,
\label{rb2cdefg}
\ee
which in the theory (IV) is consistent with 
$r_{b_2} < r_{\max} < g^2/ 64 \pi  \omega^3 = g^2/ 64 \pi \langle E \rangle^3 = 10^{11} \ell_{\rm P}$ (for $g = M_{\rm P}$) in (\ref{gmass}). 
%%%

In general for $\langle E \rangle = 10^{-\alpha} M_{\rm P}$, $n=4$ (radiation), $a \sim t^{1/2}$, $t=t_{\rm inflation} \sim 10^{-31}$ sec, $t_0 = 10^{17}$ sec, 
$\rho_0 = 1.8 \times 10^{-29} {\rm gr}/{\rm cm}^3 = 1.8 \times 10^{-24} M_{\rm P}/{\rm cm}^3$ ($M_{\rm P} = 10^{-5} {\rm gr}$),  the inequality (\ref{rb2c}) turns into:
\be
r_{b_2} \gtrsim \left[ \frac{10^{- \alpha} M_{\rm P}}{\rho_0}  \left( \frac{t_{\rm inflation}}{t_0} \right)^2 \right]^{\frac{1}{3}} \approx  4 \times 10^{-23 - \alpha/3} {\rm cm} = \,  4 \times 10^{10 - \alpha/3}  \ell_{\rm P} 
\, .
\label{rb2cdefgG}
\ee
According to (\ref{gmass}) for the theory (IV), 
$r_{b_2} < r_{\max} < g^2/ 64 \pi  \omega^3 = g^2/ 64 \pi \langle E \rangle^3 = 10^{-5+3 \alpha} \ell_{\rm P}$ (for $g = M_{\rm P}$) or $\alpha > 9/2$.

According to (\ref{constraint gs2}), for the theory (III) the condition for having bound states (\ref{rb2cdefgG}), assuming again the energy 
to be $\langle E \rangle = 10^{-\alpha} M_{\rm P}$, 
has to be consistent with
is $r_{b_2} < r_{\rm max} < e^2/8 \pi \omega = 4 \times 10^{-4 + \alpha} \ell_{\rm P}$ or  $\alpha > 21/2$ 
(consistently with the perturbative expansion we here assumed the coupling constant to be $e = 0.1$).

Summarizing, in the theory (III) we can create bound states at the end of inflation with energy $\omega = 10^{-\alpha} M_{\rm P}$ for $\alpha > 11.5$, while in the theory (IV) for $\alpha > 4.5$.

Therefore, bound states for the theories (III) and (IV) can serve as dark matter formed after inflation. 

Once again, we remark that the analysis in this paragraph is very preliminary, and it is not the main outcome of the paper. However, in order to support the idea of dark matter as bound states, such preliminary analysis seems promising.

%%%%% Bremsstrahlung %%%%%

Let us make a final comment about the Bremsstrahlung effect and the stability of the bound states. By such effect, we here mean the production of gravitational, electromagnetic, or any other kind of radiation by the deceleration of a particle when deflected by another particle. The moving particle loses energy, which is converted into radiation (i.e., gravitons, photons, etc.) according to the energy conservation. 

In quantum field theory or, more precisely, in the perturbative Feynman expansion, the Bremsstrahlung effect happens throughout the emission of gravitons or photons, etc., but at a higher-order in the coupling constant's expansion. Therefore, it is next to the leading order in the perturbative expansion implemented here, because it consists of adding extra external legs to the tree-level scattering amplitude. However, such an effect in principle exists, and the loss of energy in favor of gravitons, and/or other particles may affect the stability of the bound states. A very preliminary inspection of the results in this paper shows that electroballs, and quantum perturbative gaugeballs will survive the Bremsstrahlung emission because they require $\omega < M_{\rm P}$, while the stringballs seem unstable because their energy should be $\omega > M_{\rm P}$. Nevertheless, if the softness of the string amplitudes is preserved at quantum level, then the higher-order corrections in the coupling constant should be harmless for the stability. The same argument should be true for the case of asymptotically free theories. Therefore, on the basis of our present knowledge, we think that bound states are perturbatively stable.

%%%%%%%%%%%%%%%%%%%%%%%%%%%%%%%%%%%%%%%%%%%%%%%%%%%%%%%%%%%%%%%%%%%
%%%%%%%%%%%%%%%%%%%%%%%%%%%%%%%%%%%%%%%%%%%%%%%%%%%%%%%%%%%%%%%%%%%  

\acknowledgments
This work was supported by the Basic Research Program of the Science, Technology, and Innovation Commission of Shenzhen Municipality (grant no. JCYJ20180302174206969). 

%%%%%%%%%%%%%%%%%%%%%%%%%%%%%%%%%%%%%%%%%%%%%%%%%%%%%%%%%%%%%%%%%%%
%%%%%%%%%%%%%%%%%%%%%%%%%%%%%%%%%%%%%%%%%%%%%%%%%%%%%%%%%%%%%%%%%%%  

\appendix
\section{Extension of the bound states as a function of the impact factor $b$ and energy $\omega$} 
\label{extension}

Here we discuss how the turning points $r_{b_1}$ and $r_{b_2}$ changes with respect to the variations of $b$ or $\omega$, with the other relevant parameters fixed.

\begin{figure}[h]
\includegraphics[scale=1]{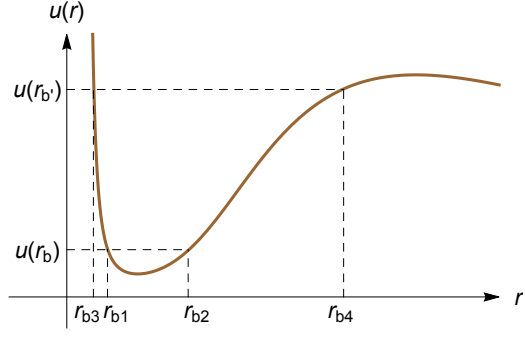}
\caption{\small {%Turning points corresponding to the functions $u(r_b') > u(r_{b})$. 
Turning points of the  function $u(r)$ for two different values of the impact parameter, namely $b' < b$ and $u(r_b') > u(r_{b})$.
As the plot shows, $ |r_{b4} - r_{b3}| > |r_{b2} - r_{b1}|$. 
When $b$ decreases, according to \eq{urb}, $u(r_b) \to u(r_{b'})$ and as consequence $ |r_{b2} - r_{b1}| \to |r_{b4} - r_{b3}| $. Thus, in this situation the bound state size increases. The opposite situation is also true. When $b$ increases, $u(r_b') \to u(r_{b})$, $ |r_{b4} - r_{b3}| \to |r_{b2} - r_{b1}| $ and the bound state size reduces.   
 }\label{u-r}}
\end{figure}

$\bullet$ We start by considering variations of $b$, with $\omega$ and $a$ fixed. Let $r_b$ denote the turning points $r_{b_1}$ or $r_{b_2}$. As discussed in Sec.~\ref{sec2}, the turning points are the solution of the equation
\beq
\label{U12}
U(r_b) = E_\text{eff} = \frac12
.
\eeq
Using (\ref{effective potential energy}) in \eq{U12}, the above equation can be rewritten as:
\be
u(r_b)=\left(1 + \frac{a^2}{b^2} \right) u(r_0)
\,,
\quad
\text{where}
\quad
u(r)=\frac{1}{r^2}\, e^{\frac{V(r)}{\omega}} 
\,.
\label{urb}
\ee
In the scattering situation, $a\gg b$ and $r_0=\sqrt{a^2+b^2}/2\approx a/2$. Therefore, $u(r_0) \approx u(a/2)$ can be treated approximately as a constant whether we keep fixed the parameters $a$ and $\om$. Thus, in this situation \eq{urb} implies that if $b$ decreases (increases) the value of the function $u(r_b)$ increases (decreases). Therefore, the size of the interval $[r_{b_1},r_{b_2}]$ gets bigger (smaller) whether $b$ decreases (increases), because $r_{b_1}$ and $r_{b_2}$ are the endpoints of a valley of $u(r)$, see Fig.~\ref{u-r}.

The largest value for $r_{b_2}$ follows from the bound state condition $U_\text{max} > E_\text{eff}$ in \eqref{nsconditions}, which implies
\be
b>a \left[\frac{u(r_\text{max})}{u(r_0)}-1\right]^{-1/2} \, . 
\label{lower bound b}
\ee
If $b$ approaches the above lower bound, \eqref{urb} turns into equation $u(r_b)=u(r_\text{max})$ with the solution $r_{b_2}=r_\text{max}$. Thus, $r_\text{max}$ is an upper bound for the extension $r_{b_2}$. The largest value of $r_\text{max}$ determines the maximum for $r_{b_2}$. 

\begin{figure}[h]
\includegraphics[scale=0.7]{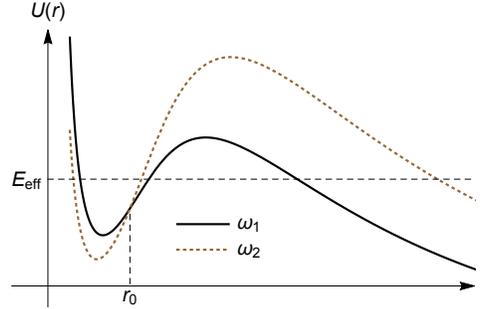}
\caption{\small Effective potential of gravi-scalarballs for different values of $\omega$ when $a$, $b$, and $\mu$ are fixed. In the plot $\omega_2 > \omega_1$. \label{Gravi-scalar omega}}
\end{figure}

$\bullet$ Now, we show how $r_{b}$ varies by increasing $\omega$, with $b$ and $a$ fixed. Let us start with the case of \underline{gravi-scalarballs}, described in Sec.~\ref{sec3}. 
Using \eq{deltaSFT} we can write the effective potential~\eq{effective potential energy} explicitly, namely, 
\be
U(r) = U_0 \left( \frac{r_0}{r}\right)^2 \exp \left\{4G\omega \mu \left[\frac{{\rm erf}(\mu r_0)}{\mu r_0}-\frac{{\rm erf}(\mu r)}{\mu r}\right]\right\},
\quad\text{where}\quad
U_0 = \frac{b^2}{2(a^2+b^2)} \, .
\label{U omega}
\ee
Because $|{\rm erf}(x)| \leqslant |x| $, the expression in the square brackets of \eq{U omega} is negative for $r < r_0$ and is positive for $r > r_0$. Thus, for $r<r_0$ the function $U(r)$ is a monotonously decreasing with respect to $\omega$, while for $r>r_0$ it is a monotonously increasing function of $\omega$. Therefore, when we increase $\omega$ the curve of $U( r)$ moves down in the region $r<r_0$ and it moves up in the region $r>r_0$, see, e.g, Fig.~\ref{Gravi-scalar omega}. As the Fig.~\ref{Gravi-scalar omega} shows, both $r_{b_1}$ and $r_{b_2}$ decreases if $\omega$ increases. Moreover, we see that $r_\text{min}$ decreases while $r_\text{max}$ increases whether $\omega$ increases. In particular, $r_\text{max}\rightarrow 2G\omega$ 
when $\omega\rightarrow\infty$.

For the {\underline{stringballs}}, we just need to replace $\mu$ by $M(\omega,\alpha')$ in \eq{U omega}. $M(\omega,\alpha')$ is given by Eq.~\eqref{M omega alpha}. Considering the bound state requirement $\omega\gg 1/\sqrt{\alpha'}$, we can regard $M(\omega,\alpha')$ as a small quantity and write down the effective potential as
\be
U(r)\approx U_0\Bigl( \frac{r_0}{r}\Bigr)^2 \exp\biggl[ \frac{8G(r^2-r_0^2)}{3\sqrt{\pi}} \, \omega M(\omega,\alpha')^3\biggr] \, .
\ee
One can prove that 
\be
\omega M(\omega,\alpha')^3
=\Biggl[\frac{2}{\alpha'^{4/3}}\cdot\frac{(\alpha'\omega^2)^{1/3}}{\log(\alpha'\omega^2)} \Biggr]^{3/2}
\ee
is a monotonously increasing function when $\omega>\sqrt{e^3/\alpha'}$ and it is a monotonously decreasing function for $\omega < \sqrt{e^3/\alpha'}$. Thus, the change of the curve $U(r)$ is similar to the case of gravi-scalarballs, shown in Fig.~\ref{Gravi-scalar omega}. Hence, we end up with the same conclusion: if $\omega$ increases, both $r_{b_1}$ and $r_{b_2}$ decrease, and $r_\text{min}$ decreases while $r_\text{max}$ increases. Finally, $r_\text{max}\rightarrow 2G\omega$ when $\omega\rightarrow\infty$.

Given the different dimension of the coupling constant of the \underline{electroballs}, and \underline{scalarballs} in fourth derivative, sixth derivative, and nonlocal theories, the conclusions for these models are opposite to the one above. If $\omega$ increases, both $r_{b_1}$ and $r_{b_2}$ increase, and $r_\text{min}$ increases while $r_\text{max}$ decreases.

%%%%%%%%%%%%%%%%%%%%%%%%%%%%%%%%%%%%%%%%%%%%%%%%%%%%%%%%%%%%%%%%%%%
%%%%%%%%%%%%%%%%%%%%%%%%%%%%%%%%%%%%%%%%%%%%%%%%%%%%%%%%%%%%%%%%%%%

%%%%%%%%%%%%%%%%%%%%%%%%%%%%%%%%%%%%%%%%%%%%%%%%%%%%%%%%%%%%%%%%%%%
%%%%%%%%%%%%%%%%%%%%%%%%%%%%%%%%%%%%%%%%%%%%%%%%%%%%%%%%%%%%%%%%%%%

\end{document}